\documentclass[a4paper,11pt]{article}
\usepackage{graphicx,amssymb,amstext,amsmath,mathabx,mathtools,esvect,xparse}
\usepackage{color}
\usepackage{floatflt}

\definecolor{cbl}{rgb}{0,0,1}               

\newcommand{\bc}{\begin{center}}
\newcommand{\ec}{\end{center}}
\def\ba#1{\begin{array}{#1}\displaystyle}
\newcommand{\ea}{\end{array}}

\newcommand{\beq}{\begin{equation}}
\newcommand{\eeq}{\end{equation}}
\newcommand{\beqa}{\begin{eqnarray}}
\newcommand{\eeqa}{\end{eqnarray}}

\newcommand{\bi}{\begin{itemize}}
\newcommand{\ei}{\end{itemize}}

\newcommand{\bra}{\langle}
\newcommand{\ket}{\rangle}

\newcommand{\TTb}{\mathrm{T}\overline{\mathrm{T}}}
\newcommand{\bal}{\boldsymbol{\alpha}}
\newcommand{\bol}{\boldsymbol{0}}
\newcommand{\bel}{\boldsymbol{\beta}}

\usepackage{cite}

\definecolor{purple_nice}{rgb}{0.4,0.2,0.7}
\definecolor{fuel_blue}{RGB}{42,162,185}
\definecolor{YInMn_blue}{RGB}{46, 80, 144}
\definecolor{ultramarine}{RGB}{63, 0, 255}
\definecolor{KLEIN_blue}{rgb}{0, 0.18, 0.65}
\usepackage[linktocpage=true]{hyperref}
\hypersetup{
    colorlinks=true,
    linkcolor=YInMn_blue,
    citecolor=ultramarine,
    filecolor=fuel_blue,
    urlcolor=KLEIN_blue,
}

\makeatletter
\renewenvironment{abstract}{%
      \begin{center}%
        {\bfseries \normalsize\abstractname\vspace{\z@}}
      \end{center}%
      \quotation}
    {\endquotation}
\makeatother
\usepackage[normalem]{ulem}  

\begin{document}

\begin{titlepage}
\title{On the Representation of Minimal Form Factors\\ in Integrable Quantum Field Theory}
\author{Olalla A. Castro-Alvaredo${}^{\heartsuit}$, Stefano Negro${}^\clubsuit$ and Istv\'an M. Sz\'ecs\'enyi${}^{\diamondsuit,\spadesuit}$\\[0.3cm]}
\date{\small ${}^{\heartsuit}$ Department of Mathematics, City, University of London,\\ 10 Northampton Square EC1V 0HB, UK\\[0.3cm]
${}^{\clubsuit}$ C.N. Yang Institute for Theoretical Physics, State University of New York\\ Stony Brook, NY 11794-3840. U.S.A.\\
[0.3cm]
${}^{\diamondsuit}$ Nordita, KTH Royal Institute of Technology and Stockholm University,\\ Hannes Alfv\'ens v\"ag 12, SE-106 91 Stockholm, Sweden\\
[0.3cm]
${}^{\spadesuit}$ University of Modena and Reggio Emilia, FIM,\\ via G. Campi 213/b, 41125, Modena, Italy\\
}
\maketitle
\begin{abstract}
In this paper, we propose a new representation of the minimal form factors in integrable quantum field theories. These are solutions of the two-particle form factor equations, which have no poles on the physical sheet. Their expression constitutes the starting point for deriving higher particle form factors and, from these, the correlation functions of the theory. As such, minimal form factors are essential elements in the analysis of integrable quantum field theories. The proposed new representation arises from our recent study of form factors in $\TTb$-perturbed theories, where we showed that the minimal form factors decompose into elementary building blocks. Here, focusing on the paradigmatic sinh-Gordon model, we explicitly express the standard integral representation of the minimal form factor as a combination of infinitely many elementary terms, each representing the minimal form factor of a generalised $\TTb$ perturbation of the free fermion. Our results can be readily extended to other integrable quantum field theories and open various relevant questions and discussions, from the efficiency of numerical methods in evaluating correlation functions to the foundational question of what constitutes a ``reasonable'' choice for the minimal form factor.
\end{abstract}
\vfill

\noindent 
${}^{\heartsuit}$ o.castro-alvaredo@city.ac.uk\\
${}^{\clubsuit}$ stefano.negro@stonybrook.edu\\
${}^{\diamondsuit,\spadesuit}$ szistvan87@gmail.com
\hfill \today
\end{titlepage}

\section{Introduction}
In 2016, Smirnov and Zamolochikov published an investigation of the ``space of integrable quantum field theories" \cite{Smirnov:2016lqw}. They showed that for every integrable quantum field theory (IQFT), with its associated factorised scattering ($S$) matrix (that would include the sine-Gordon, sinh-Gordon and Ising models, for instance), we can construct an infinite family of IQFTs whose $S$-matrix differs from the original by CDD factors \cite{Castillejo:1955ed}\footnote{What this means is that the $S$-matrix of the new theory is the product of the original $S$-matrix times factors which trivially satisfy all physical requirements (Lorenz invariance, unitarity, hermitian/real analyticity, etc.) but do not introduce any new poles, that is, they do not change the particle spectrum of the model.}. The fact that the $S$-matrix bootstrap equations admit a CDD ambiguity was not a new result, but the interpretation and form of these CDD factors were. They showed that the space of all CDD factors is spanned by exponentials of hyperbolic functions, with exponents proportional to the one-particle eigenvalues of the local conserved quantities of the IQFT (of which there are infinitely many because of integrability). They showed that each of these exponential factors, in turn, can be interpreted as arising from a perturbation of the action of the original IQFT by an irrelevant operator. The simplest of these operators is the $\TTb$ field, the composition of the holomorphic and anti-holomorphic components of the stress-energy tensor. If the original theory has a single-particle spectrum with a two-body scattering matrix $S_{\bol}(\theta)$, where $\theta$ is the rapidity difference between two scattering particles, then in the perturbed model, the scattering matrix would be 
\beq 
S_{\bal}(\theta):=S_{\bol}(\theta) \Phi_{\bal}(\theta)\,, \quad \mathrm{with}\quad \log(\Phi_{\bal}(\theta))=-i\sum_{s\in \mathcal{S}} \alpha_s m^{2s} \sinh(s\theta)\,,
\label{Smatrix}
\eeq 
where $\bal:=\{\alpha_s\}_{s\in \mathcal{S}}$ is a set of coupling constants, $m$ is a mass scale, and $s$ are integers taking values in the set $\mathcal{S}$. $\mathcal{S}$ is typically the set of spins of local conserved quantities. Thus, for every choice of the set $\bal$, there is a distinct integrable $S$-matrix. 

Following this result, $\TTb$-perturbed theories (corresponding to having only one term in the sum, i.e. $s=1$) and their generalisations have been extensively studied from many different viewpoints. In fact, there were already a few interesting related results, even before \cite{Smirnov:2016lqw}. One of the first studies of the operator $\TTb$ was presented in \cite{Zamolodchikov:2004ce}, while  \cite{sgMuss} investigated the correlation functions of the bosonic version of the sinh-Gordon model, showing they display certain properties related to the interpretation of the model as sinh-Gordon perturbed by infinitely many irrelevant operators. Following these works, an analysis of the form factors of the operator $\TTb$ in unperturbed theories appeared in \cite{Delfino:2004vc, Delfino:2006te}. 

The famous thermodynamic Bethe ansatz (TBA) study carried out in \cite{Cavaglia:2016oda} provided a new viewpoint on the infrared (IR) and ultraviolet (UV) physics of these models. For a single $\TTb$ perturbation,  the properties of the theory are fundamentally different depending on the sign of the coupling $\alpha:=\alpha_1$. In particular, for $\alpha<0$, a so-called Hagedorn transition takes place, which is associated with a square-root branch point appearing in the free energy at a finite, positive value of the compactification radius. In addition, the TBA equations dictate that while the perturbation does not change the IR properties of the model, it profoundly alters its UV behaviour. In fact, it is no longer a quantum or conformal field theory in the standard sense. This fact is sometimes described as a ``lack of UV completion" or ``UV fragility'' and has been discussed in various places \cite{Dubovsky:2012wk, Dubovsky:2013ira, Dubovsky:2017cnj}. The TBA approach allowed the analysis of several generalisations of the $\TTb$ deformation\footnote{An interesting generalisation is the so-called \emph{root $\TTb$} deformation \cite{Rodriguez:2021tcz,Babaei-Aghbolagh:2022uij,Ferko:2022cix,Babaei-Aghbolagh:2022leo,Garcia:2022wad,Borsato:2022tmu} which is intimately related to the Modified Maxwell (ModMax) theory \cite{Bandos:2020jsw,Kosyakov:2020wxv,Bandos:2020hgy,Sorokin:2021tge}.} \cite{Conti:2019dxg, Hernandez-Chifflet:2019sua, Camilo:2021gro, Cordova:2021fnr, LeClair:2021opx, LeClair:2021wfd, Ahn:2022pia}. These have also been studied by other methods such as perturbed conformal field theory \cite{Guica:2017lia, Cardy:2018sdv, Cardy:2019qao, Aharony:2018vux, Aharony:2018bad, Guica:2020uhm, Guica:2021pzy, Guica:2022gts},  string theory \cite{Baggio:2018gct, Dei:2018jyj, Chakraborty:2019mdf, Callebaut:2019omt}, holography \cite{McGough:2016lol, Giveon:2017nie, Gorbenko:2018oov, Kraus:2018xrn, Hartman:2018tkw, Guica:2019nzm, Jiang:2019tcq,Jafari:2019qns}, quantum gravity \cite{Dubovsky:2017cnj, Dubovsky:2018bmo, Tolley:2019nmm, Iliesiu:2020zld, Okumura:2020dzb, Ebert:2022ehb}, out-of-equilibrium conformal field theory \cite{Medenjak:2020ppv, Medenjak:2020bpe}, long-range spin chains \cite{Bargheer:2008jt,Bargheer:2009xy,PJG, Marchetto:2019yyt}, and the generalised hydrodynamics (GHD) approach \cite{Doyon:2021tzy,Cardy:2020olv,Doyon:2023bvo}. The latter treatment provides an alternative interpretation of the special features of generalised $\TTb$ perturbations. In particular, for $\TTb$, we may see the CDD factor in the $S$-matrix as a phase factor that induces a particle length. Thus, the fundamental degrees of freedom become extended with positive or negative length, depending on the sign of $\alpha$, which provides a new interpretation of the UV physics of these models. An important contribution of \cite{Doyon:2021tzy} is the realisation that the CDD factor above can be further generalised to include the full set of extensive conserved charges of the IQFT. A generalised perturbation in the above sense can then give rise to the $S$-matrices of any IQFT, hence achieving the original aspiration of \cite{Smirnov:2016lqw} of truly constructing ``the space of IQFTs". 
This result goes hand in hand with a much better understanding of the critical role played by non-local charges in the dynamics of out-of-equilibrium integrable models (see e.g.~\cite{enej}).
This means, in particular, that the CDD factors -- with some abuse of nomenclature -- can also be generalised to introduce new bound states, in which case the spectrum of the original unperturbed theory needs to be consistently completed, as discussed in \cite{Doyon:2021tzy}.

Recently, in a series of papers involving two of the present authors \cite{PRL,longpaper,entropyTTB} (see also \cite{Hou}), a study of the operator content of theories with $S$-matrix (\ref{Smatrix}) was carried out. We employed the form factor program \cite{KW,smirnovBook}, a standard approach to computing the matrix elements of local and semi-local fields in IQFT, and applied it to generalised $\TTb$-perturbed theories. We found solutions for the form factors of various models and fields, establishing that, like the $S$-matrix, the form factors of perturbed theories factorise into the unperturbed form factors and a function of $\bal$. We found that the form factors may depend on additional arbitrary constants $\bel=\{\beta_s\}_{s\in \mathcal{S}'}$ that arise in the solution to the minimal form factor (MFF) equations. MFFs are a fundamental building block of all form factor solutions.  For the $S$-matrix (\ref{Smatrix}), there is a single MFF, which satisfies the equations 
\beq 
f(\theta;\bal)=S_{\bal}(\theta) f(-\theta;\bal)=f(-\theta+2\pi i;\bal)\,.
\label{minieq}
\eeq 
In \cite{PRL,longpaper}, we showed that the most general solution to these equations is
\beq 
f(\theta;\bal)=f(\theta;\bol) \varphi_{\bal}(\theta)C_{\bel}(\theta)\,,
\label{mineq}
\eeq 
with $f(\theta;\bol)$ the MFF of the unperturbed theory, while\footnote{The formula for $\varphi_{\bal}(\theta)$ was, to our knowledge, first derived by I.M.~Sz\'ecs\'enyi in an unpublished work, while its massless version appeared in an investigation in Nambu-Goto theory \cite{Dubovsky:2012wk}, which later turned out to be the $\TTb$ deformation of a free, massless, 2-dimensional scalar field theory.}
\beq 
\log(\varphi_{\bal}(\theta))=\frac{ \theta-i\pi}{2\pi} i \log(\Phi_{\bal}(\theta))\,,\quad \mathrm{and} \quad \log(C_{\bel}(\theta))=\sum_{s\in \mathcal{S}'} \beta_s m^{2s} \cosh(s\theta)\,.
\label{mini}
\eeq 
As mentioned above, an interesting property of this solution is the presence of free parameters $\bel$, which are absent in the $S$-matrix. Whereas the function $\varphi_{\bal}(\theta)$ is essential in solving (\ref{mineq}), the function $C_{\bel}(\theta)$ is a sort of CDD factor of the MFF, in the sense that any solution $\varphi_{\bal}(\theta)$ can be multiplied by a function $C_{\bel}(\theta)$ for any choice of parameters $\bel$ and any choice of integers $s \in \mathcal{S}'$ and still give a solution to (\ref{mineq}). The presence of factors $C_{\bel}(\theta)$ changes the asymptotic properties of the MFF, hence also of higher particle form factors, where products of $f(\theta;\bal)$ are typically involved.  According to the usual interpretation that each distinct solution to the form factor equations corresponds to a different local or semi-local field, it is tempting to interpret each choice of $\bel$, that is, each choice of the MFF, as the choice of a different building block for the form factors of local and semi-local fields. In other words, the couplings $\bel$ parametrise some kind of background feature of the theory itself, which couples to and alters all local and semi-local fields. However, at the moment, we do not possess enough evidence to support this interpretation, so we refrain from making a definitive statement. We believe this question to be important and plan to investigate it in future works.

In the analysis of \cite{PRL,longpaper}, the choice $\bel=\bol$ was made. Since most of the work focused on $\TTb$-perturbed theories, only a single non-zero parameter appeared in (\ref{mini}). In this paper, we will see that, while the role of the parameters $\bel$ in the form factors of generalised $\TTb$-perturbed theories is not entirely clear, the functions $C_{\bel}(\theta)$ play a fundamental part in the computation of the MFFs of more standard IQFTs. 

The works \cite{Hou,entropyTTB} extended the study of \cite{PRL,longpaper}  to a class of fields called branch point twist fields (BPTFs). These fields are important in the context of entanglement measures \cite{entropy}; therefore, their form factors are also of interest. BPTFs emerge in replica theories, that is, models consisting of $n$ identical copies of a given IQFT and where a large amount of symmetry is present. BPTFs are associated with cyclic permutation symmetry, and their form factors satisfy the generalised equations proposed in \cite{entropy}. The solution procedure also centres around an MFF which, for generalised $\TTb$-perturbed theories and particles in the same copy, satisfies a slightly modified version of (\ref{mini}), that is 
\beq 
f_n(\theta;\bal)=S_{\bal}(\theta) f_n(-\theta;\bal)=f_n(-\theta+2\pi i n;\bal)\,.
\label{minieq2}
\eeq 
The most general solution to this equation was simultaneously found in \cite{Hou,entropyTTB} and has a very similar structure to that discussed earlier, except that solutions now depend also on $n$:
\beq 
f_n(\theta;\bal)=f_n(\theta;\bol) \varphi_{\bal}^n(\theta)C^n_{\bel}(\theta)\,,
\label{mineq2}
\eeq 
with $f_n(\theta;\bol)$ the MFF of the unperturbed theory with  $f_1(\theta;\bal):=f(\theta;\bal)$, and 
\beq 
\log(\varphi_{\bal}^n(\theta))=\frac{ \theta-i\pi n}{2\pi n} i \log(\Phi_{\bal}(\theta))\,,\quad \mathrm{and} \quad \log(C_{\bel}^n(\theta))=\sum_{s\in \mathcal{S}'} \beta_s m^{\frac{2s}{n}}  \cosh \frac{s\theta}{n}\,.
\label{mini2}
\eeq 
Even if the representation (\ref{Smatrix}) suggests that the scattering matrices of generalised $\TTb$-perturbed IQFTs are fundamentally different from those of ``standard" IQFTs, the fact is that there is no difference between the two sets provided we allow the sum over spins in (\ref{Smatrix}) to contain infinitely many terms. Stated differently, any factorisable $S$-matrix admits a representation (\ref{Smatrix}) (or a suitable generalisation for a multi-particle spectrum), provided the appropriate choice of the parameters $\alpha_s$ is made. In many cases, this choice involves a sum over infinitely many parameters, in which case the CDD factor $\Phi_{\bal}(\theta)$ typically becomes a product of ratios of hyperbolic functions of the rapidities \cite{Camilo:2021gro}. 

One way to think about the above statement is to consider the renowned sinh-Gordon model, which we will use as a representative example in this paper. The $S$-matrix of the sinh-Gordon model is well known to be a simple ratio of hyperbolic tangent functions of the rapidity and the sinh-Gordon coupling \cite{SSG2,SSG3}. At the same time, it is clear that the sinh-Gordon $S$-matrix is a CDD factor that is a solution to all consistency equations for the $S$-matrix, which has no poles on the physical strip in rapidity space. It follows, then, that the sinh-Gordon $S$-matrix must admit a representation of the form $\Phi_{\bal}(\theta)$ for particular values of $\alpha_s$, which, as we show below, it does.  Therefore, we expect the MFF of this theory for local, semi-local and BPTFs to be of the form (\ref{mini})-(\ref{mini2}), representations that are very different from the usual integral ones first proposed in \cite{KW} (and in \cite{entropy} for the BPTF). Here, we will show that these two representations are truly equivalent, provided the functions $C_{\bel}(\theta), C_{\bel}^n(\theta)$ take a specific, non-trivial form. Our results for sinh-Gordon can be easily extended to other IQFTs, especially to those which are very directly related to sinh-Gordon, such as the Lee-Yang model \cite{CMLeeYang} and the first breather sector of the sine-Gordon model \cite{Zamolodchikov:1977yy,Zamolodchikov:1978xm}. 

\medskip

We organised this paper as follows. Section \ref{sec:2} reviews some facts about the structure of the $S$-matrices and minimal form factors of IQFTs, focusing on their integral representations. In Section \ref{sec:3}, we introduce the $S$-matrix of the sinh-Gordon model and recast it as a generalised $\TTb$-perturbed Ising field theory. In Section \ref{sec:4}, we do the same for the MFF of the sinh-Gordon model for the case of local and semi-local fields. We generalise to the BPTF in Section \ref{sec:minff_BPTF}. Section \ref{sec:MFF_properties} discusses the analyticity,  asymptotics and other properties of the MFF.  We conclude in Section \ref{sec:7}. We collect the technical derivations in Appendices~\ref{ApA} to \ref{numerics}.

\section{Building Blocks and Integral Representations: a Brief History}\label{sec:2}
It is well known that the scattering amplitudes of diagonal IQFTs -- that is, theories where there is no back-scattering -- can be expressed in terms of blocks of the form \cite{exact,mussardobook}
\beq
(x)_\theta:=\frac{\sinh\frac{1}{2}\left(\theta+ i\pi x\right)}{\sinh\frac{1}{2}\left(\theta- i\pi x\right)}\,.
\label{block}
\eeq
Such blocks are extremely natural since they easily allow for the construction of two-body scattering amplitudes $S_{ab}(\theta)$ satisfying the properties of unitary and crossing
\beq
S_{ab}(\theta)S_{ba}(-\theta)=1\,, \qquad S_{ab}(i\pi-\theta)=S_{b\bar{a}}(\theta)=1\,.
\label{uc}
\eeq 
Here $S_{ab}(\theta)$ is the scattering amplitude of the process $a+b \mapsto a+b$,  and $\bar{a}$ is the particle conjugate to $a$. In many cases, such as when there is a single particle in the spectrum ($S_{ab}(\theta):=S(\theta)$), or when there is parity invariance ($S_{ab}(\theta)=S_{ba}(\theta)$), and particles are self-conjugate ($\bar{a}=a$), all functions in (\ref{uc}) are the same. Then, it is immediate to see that the following particular block combination
\beq 
[x]_\theta:=-(x)_\theta (1-x)_\theta= \frac{\tanh\frac{1}{2}\left(\theta+ i\pi x\right)}{\tanh\frac{1}{2}\left(\theta- i\pi x\right)}\,,
\label{blocksq}
\eeq 
satisfies the relations
\beq 
[x]_\theta [x]_{-\theta}=1 \quad \mathrm{and} \quad [x]_{\theta}=[x]_{i\pi-\theta}\,.
\eeq 
This implies that any function consisting of products of these blocks will automatically satisfy unitarity and crossing. These are, however, not the only constraints when constructing the scattering matrices of an IQFT. The spectrum of the theory and, especially, its bound state structure play a critical role in deciding which specific blocks -- $(x)_\theta$ or $[x]_\theta$ -- and which values of $x$ are present in which scattering amplitude. One usually imposes these constraints via the $S$-matrix bootstrap equations. These will play no role in our work, so we will not discuss them here. As examples of the above, let us mention that the sinh-Gordon model \cite{SSG2,SSG3} and Lee-Yang model \cite{CMLeeYang} both have $S$-matrices consisting of just a single block $[x]_\theta$ (in the Lee-Yang case this block also has a physical pole), whereas the family of minimal and affine Toda field theories generally have complicated multi-particle spectra but still admit $S$-matrix representations \cite{BCDS,FKS} built out of the blocks above. Even for non-diagonal theories, such as the sine-Gordon model, the blocks \eqref{block} and \eqref{blocksq} are useful. In sine-Gordon, there is a diagonal part of the spectrum (the breather sector) where the $S$-matrices can be expressed in terms of them \cite{Zamolodchikov:1977yy,Zamolodchikov:1978xm}. Even for the non-diagonal sector, the scattering amplitudes admit integral representations of the same type as those that can be written for $(x)_\theta$ and $[x]_\theta$. This naturally brings us to the topic of integral representations. 

It is easy to show that\footnote{The minus sign can be absorbed into the integral representation by employing the formal relation 
\beq
-1=\exp\left(\pm 2\int_0^\infty \frac{dt}{t} \sinh\frac{t\theta}{i\pi}\right)\,,
\eeq 
as we have done in (\ref{xsqu}).} 
\beq 
(x)_\theta= -\exp\left(2 \int_0^\infty \frac{dt}{t} \frac{\sinh t(1-x)}{\sinh t} \sinh\frac{t\theta}{i\pi}\right)\,,
\label{xcur}
\eeq 
for $0<x<1$\footnote{For $-1<x<0$ we can use the equality $(x)_\theta=(-x)_\theta^{-1}$, thus we have the representation 
\beq 
(x)_\theta= -\exp\left(-2 \int_0^\infty \frac{dt}{t} \frac{\sinh t(1+x)}{\sinh t} \sinh\frac{t\theta}{i\pi}\right)\,,
\label{xcurnew}
\eeq
instead. This is in fact the representation that is required in the sinh-Gordon case where the $S$-matrix is the block $[-B/2]_\theta$ with $0<B<2$.} and, similarly,
\beq 
[x]_\theta= \exp\left(-8 \int_0^\infty \frac{dt}{t} \frac{\sinh \frac{t(1-x)}{2} \sinh \frac{t x}{2} \sinh\frac{t}{2}}{\sinh t} \sinh\frac{t\theta}{i\pi}\right)\,.
\label{xsqu}
\eeq 
It follows that all diagonal scattering matrices admit a representation in terms of products of the integral blocks above. 
The integral representations are helpful in many contexts. For instance, they allow for a universal representation of all the $S$-matrices of affine Toda field theory \cite{FKS}. These are, in turn, very useful in the context of TBA and in Toda theories, to an algebraic structure which links the particle content of the theory to the Dynkin diagram of a finite Lie algebra \cite{tba2,dynkin0,dynkin1,fringTBA,ourTBA}.

Another context where an integral representation of the $S$-matrix is useful is when solving the form factor equations (\ref{minieq}) and (\ref{minieq2}). It was shown in \cite{KW} that the MFF associated with an $S$-matrix of the form 
\beq 
S(\theta)=\exp\left(\int_0^\infty \frac{dt}{t} g(t) \sinh\frac{t\theta}{i\pi}\right)\,,
\label{S}
\eeq 
admits the representation 
\beq 
f(\vartheta)=\mathcal{N}\exp\left(\int_0^\infty \frac{dt}{t} \frac{g(t)}{\sinh t} \sin^2\left(\frac{t\vartheta}{2\pi}\right)\right)\,,
\label{factor3}
\eeq 
where $\vartheta=i\pi-\theta$ and $\mathcal{N}$ is a normalisation constant. 
Here, we use the notation $f(\vartheta)$ for the MFF. Comparing (\ref{S}) to (\ref{xcur}) and (\ref{xsqu}), it is clear that each $S$-matrix block will, in turn, contribute a factor of the type (\ref{factor3}) to the minimal form factor. 

The same ideas can be applied to the minimal form factor of BPTFs, as shown in \cite{entropy}. In this case, the MFF is
\beq 
f_n(\vartheta)=\mathcal{N}\exp\left(\int_0^\infty \frac{dt}{t} \frac{g(t)}{\sinh (n t)} \sin^2\left(\frac{t\vartheta}{2\pi}\right)\right)\,,
\label{factor2}
\eeq 
with $\vartheta=i\pi n-\theta$. As expected, this reduces to $f(\vartheta)$ for $n=1$. By construction,  $\mathcal{N}=f_n(0)$. Since 
$
2\sin^2\left(\frac{t\vartheta}{2\pi}\right)=-\cos\frac{t\vartheta}{\pi}+1
$,
it is convenient to choose 
\beq 
\mathcal{N}=\exp\left(-\frac{1}{2}\int_0^\infty \frac{dt}{t} \frac{g(t)}{\sinh(nt)}\right)\,,
\eeq 
so that 
\beq 
f_n(\vartheta)=\exp\left(-\frac{1}{2}\int_0^\infty \frac{dt}{t} \frac{g(t)}{\sinh(nt)} \cos\frac{t\vartheta}{\pi}\right)\,,\quad \mathrm{and}\quad f_1(\vartheta):=f(\vartheta)\,. \label{finalMFF}
\eeq 
These MFFs look a priori quite different from the solutions (\ref{mini}) and (\ref{mini2}). However, since the blocks $(x)_\theta$, $[x]_\theta$ can also be interpreted as CDD factors, $S$-matrices made out of these blocks must be related to the deformation $\Phi_{\bal}(\theta)$ (\ref{Smatrix}). Consequently, the MFFs must admit a representation in the form of (\ref{mini}) and (\ref{mini2}). 
In the next section, we will see that this is indeed the case by focusing on the sinh-Gordon model. 

We end this section by recalling that, in the literature,  different MFF representations are used besides (\ref{finalMFF}). A good review of {relevant} identities in this regard is the Appendix of \cite{BFKZ}. A widely used representation is obtained by employing the integral representation of the $\Gamma$-function to express the formula above as an infinite product of $\Gamma$-functions. For the sinh-Gordon model, such a representation can be found, for example, in \cite{FMS}, and its generalisation to BPTFs was given in \cite{entropy}. In \cite{FMS}, a mixed representation containing an integral part and a $\Gamma$-function part was also employed and motivated by the higher efficacy of its numerical evaluation. A mixed representation of the minimal form factor of BPTFs is presented in the Appendix of \cite{Castro-Alvaredo:2020mzq}, where the focus was on the minimal $E_8$ Toda theory. However, as far as we know, a representation of the type (\ref{mini}) or (\ref{mini2}) has never been investigated.  The main result of this paper is the derivation of such a representation and a detailed study of its asymptotic and analyticity properties.

\section{The sinh-Gordon Model $S$-matrix as a CDD Factor}\label{sec:3}

\begin{figure}[h!]
\begin{center}
	\includegraphics[width=12cm]{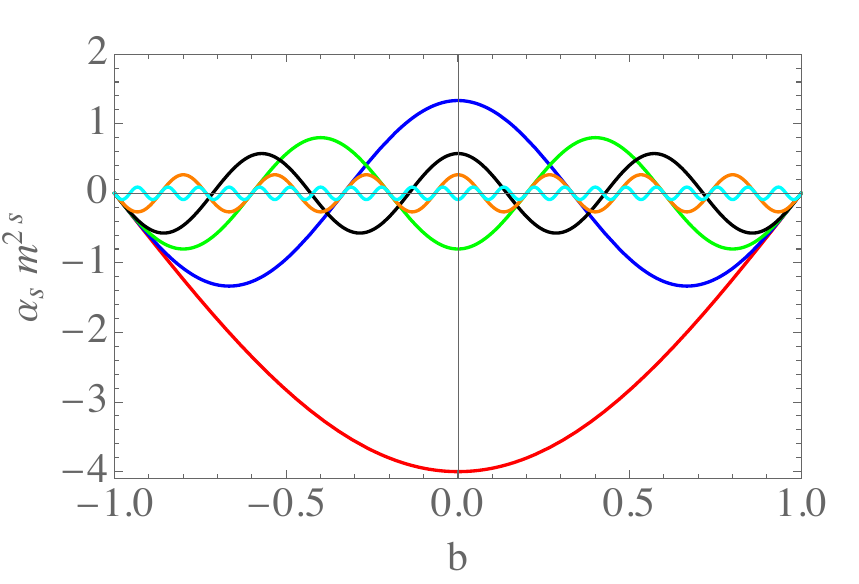}
				    \caption{The coefficients $\alpha_s m^{2s}$ in the sinh-Gordon model for $s=1,3,5,7,15,45$. The higher the value of $s$, the larger the number of oscillations. Except for $s=1$ }
				     \label{f1}
    \end{center}
\end{figure}

The sinh-Gordon $S$-matrix is simply  the block
$[-B/2]_\theta=[B/2]_\theta^{-1}$\,, 
where $B$ is the sinh-Gordon coupling, which takes values between $0$ and 2. The $S$-matrix is invariant under the exchange $B \mapsto 2-B$, a general property of affine Toda field theories (of which sinh-Gordon is the simplest example) known as weak-strong duality. To simplify the following computations, we will use the ``shifted" coupling $b=B-1$ so that the so-called self-dual point corresponds to $B=1, b=0$. In terms of $b$, the $S$-matrix can be written as
\beq 
S(\theta)= \left[-\frac{b+1}{2}\right]_\theta = \frac{\sinh\theta-i\cos\frac{\pi b}{2}}{\sinh\theta+i\cos\frac{\pi b}{2}}\,.
\label{eq:S_matrix}
\eeq 
One way to find a generalised $\TTb$ representation of this $S$-matrix is to start by finding an expansion for its logarithmic derivative
\beq 
-i \frac{d \log S(\theta)}{d\theta}=\frac{4\cos\frac{\pi b}{2} \cosh\theta}{\cos(\pi b)+\cosh(2\theta)}\,,
\label{kernel}
\eeq 
and then integrate the result in $\theta$. We note that (\ref{kernel}) admits an expansion in powers of $e^{-|\theta|}$ for $|\theta|$ large, 
which is easily inferred from the first few terms
\beq
    -i \frac{d}{d\theta}\log S(\theta) = 4 \cos\frac{\pi b}{2} e^{-|\theta|} - 4 \cos\frac{3\pi b}{2} e^{-3|\theta|} + 4 \cos\frac{5\pi b}{2} e^{-5|\theta|} + \mathcal{O}\left(e^{-7|\theta|}\right)\,.
\eeq 
The pattern is clear, and to revert to an expression which is smooth and symmetric under $\theta\to-\theta$, we can simply replace $e^{-(2k+1)|\theta|}$ with $\cosh(2k+1)\theta$, and propose the ansatz
$$
    -i \frac{d}{d\theta}\log S(\theta) = 4 \sum_{k=0}^{\infty} (-1)^k \cos\frac{(2k+1)\pi b}{2}\cosh((2k+1)\theta)\,.
$$
Integrating in $\theta$, multiplying by $i$ and exponentiating, we find the representation
\beq 
    S(\theta) = -\exp\left[-4 i \sum_{k=0}^{\infty} (-1)^{k+1}\frac{\cos\frac{(2k+1)\pi b}{2}}{2k+1} \sinh((2k+1)\theta) \right]\;.
    \label{funnyS}
\eeq 
Note that the overall minus sign comes from the integration constant, which we fix by requiring $S(0)=-1$.
Comparing to the $S$-matrix (\ref{Smatrix}) we find that the sinh-Gordon $S$-matrix is exactly of the type
\beq 
S(\theta)=S_{\bol}(\theta) \Phi_{\bal}(\theta)\,,\quad \mathrm{with}\quad S_{\bol}(\theta)=-1, 
\eeq 
with the non-vanishing couplings fixed to
\beq 
\alpha_{s} m^{2s}= \frac{4 i^{s+1}}{s} \,{\cos \frac{s b \pi}{2}}\,,\quad \mathrm{with} \quad s\in 2\mathbb{Z}^+ -1\,.
\label{alphas}
\eeq
We plotted them against $b$ for a few values of $s$ in Fig.~\ref{f1}.

Therefore, the sinh-Gordon $S$-matrix can be seen as that of the Ising field theory deformed by infinitely many irrelevant perturbations. The deep connection between the two theories is, however, not new. For instance, in \cite{Muss1}, the Ising model was recovered as a limit of the staircase model \cite{roaming}, related to sinh-Gordon by allowing $b$ to take imaginary values. 

We can write
\beq 
S(\theta)=-\Phi^{\rm shG}_{\bal}(\theta)\,, \quad \mathrm{with} \quad \bal:=\left\{\frac{4 i^{s+1}}{s\, m^{2s}} \,{\cos \frac{s b \pi}{2}}\right\}_{s\in 2\mathbb{Z}^+ -1}\,,
\eeq
where we introduced the superscript `shG' to distinguish this CDD factor from the general one. The sinh-Gordon model also provides an instructive example 
of how the sum $\log(\Phi_{\bal}(\theta))$ can have radically different analytical properties when it runs over an infinite number of terms. Whereas a single irrelevant perturbation (say $\TTb$) gives rise to $S$-matrices that are, in a sense, abnormal -- for instance, having very unusual asymptotic properties in rapidity space -- an infinite number of such perturbations can give rise to $S$-matrices which have nice physical properties and result from the (relevant) perturbation of a UV critical point\footnote{We wish to stress here that there is no simple and direct causal relation between the asymptotic properties of an $S$-matrix and the (non-)existence of a UV fixed point. In fact, as shown in \cite{Camilo:2021gro, Cordova:2021fnr}, there exist models whose $S$-matrices display a ``normal'' asymptotic behaviour -- akin to, say, sinh-Gordon or affine Toda field theories -- and nonetheless fail to have a usual UV fixed point. Two particularly simple examples are models with $S$-matrices  $S(\theta) =-[-B/2]_\theta$ and $S(\theta) = [-B/2]_\theta [-B'/2]_\theta$, with $B,B'\in(0,2)$.}. It is also worth pointing out that this analysis is easy to generalise to other IQFTs, in particular for the Lee-Yang model, whose $S$-matrix can be obtained from that sinh-Gordon by allowing $|b|$ to be larger than $1$ (in fact $b=-\frac{5}{3}$). 

\section{A New Look at the sinh-Gordon Minimal Form Factor}\label{sec:4}

Now that we have seen that the sinh-Gordon $S$-matrix can be formally written as (\ref{funnyS}), the results of our works \cite{PRL,longpaper} tell us that the MFF of the sinh-Gordon model must admit a representation of the type (\ref{mini}) for the values of $\alpha_s m^{2s}$ given in (\ref{alphas}) and some choice of the parameters $\bel$. Here, we show how to derive this representation. Using \eqref{finalMFF}, \eqref{xsqu}, and \eqref{eq:S_matrix}, we can write the logarithm of the MFF of the sinh-Gordon model as
\beqa
    \omega(\vartheta) := \log f(\vartheta) &=& - 4 \intop_{0}^\infty \frac{dt}{t} \frac{\sinh\left(\frac{1+b}{4} t\right) \sinh\left(\frac{1-b}{4} t\right) \sinh\frac{t}{2}}{\sinh^2 t}\cos\frac{\vartheta t}{\pi}\nonumber\\
    &=& -  \intop_{0}^\infty \frac{dt}{t} \frac{\cos\frac{\vartheta t}{\pi}}{\sinh t}+\intop_{0}^\infty \frac{dt}{t} \frac{\sinh\left(\frac{1+b}{2} t\right)}{\sinh^2 t}\cos\frac{\vartheta t}{\pi}+  \intop_{0}^\infty \frac{dt}{t} \frac{\sinh\left(\frac{1-b}{2} t\right) }{\sinh^2 t}\cos\frac{\vartheta t}{\pi}\,.
    \label{logarit}
\eeqa
Recall that $\vartheta = i\pi - \theta$. 
Differentiating w.r.t. $\vartheta$, we obtain
\beq
    \omega'(\vartheta) = \frac{g(\vartheta) - h(\vartheta; b) - h(\vartheta; -b)}{\pi}\,,
    \label{28}
\eeq
where
\beq
    g(\vartheta) = \intop_0^{\infty} dt\,\frac{\sin\frac{\vartheta t}{\pi} }{\sinh t}\;,\qquad h(\vartheta; b) = \intop_0^{\infty} dt\, \frac{\sinh\left(\frac{1+b}{2}t\right)}{\sinh^2 t} \sin\frac{\vartheta t}{\pi} \,.
    \label{29}
\eeq
Note that when reconstructing $\omega(\vartheta)$ from its derivative, we need to take care of an integration constant. We will fix it below.

The evaluation of the integrals \eqref{29} is straightforward and yields
\beq
    g(\vartheta) = \frac{\pi}{2} \tanh\frac{\vartheta}{2}\,,\qquad h(\vartheta;b) = \frac{1}{4} \frac{(b+1) \pi \sinh\vartheta - 2\vartheta \cos\frac{\pi b}{2}}{\cosh\vartheta + \sin\frac{\pi b}{2}}\,.
    \label{30}
\eeq
These expressions can be checked by numerical integration\footnote{In practice, the integrals are oscillatory, so it is best to use a cutoff $\Lambda\gg 1$. Note also that these results are, in principle, valid for $\left\vert\mathrm{Im}(\theta)\right\vert < \pi$. However, we can consider $\mathrm{Im}(\theta) = \pi$ as a limiting value. We will, in any case, check the final expression, so this can be considered as some kind of ansatz, to be verified \emph{ex-post}.} or by setting $n=1$ in the results of Section \ref{sec:minff_BPTF}, cf. \eqref{eq:h_n}. It is also useful to point out that the same kind of integrals are involved when we consider the contributions of individual blocks $(x)_\theta$, so these results are widely applicable. {We elaborate on this} in more detail in Appendix~\ref{apendixC}. Thus we have
\beq
    \omega'(\vartheta) = \frac{\tanh\frac{\vartheta}{2}}{2} - \frac{1}{\pi} \frac{\pi\left(\cosh\vartheta - b \sin\frac{\pi b}{2} \right) \sinh\vartheta - 2 \vartheta \cosh\vartheta \cos\frac{\pi b}{2}}{\cosh 2\vartheta + \cos \pi b}\;.
    \label{31}
\eeq
Looking at the expression above, we can already see a contribution involving the function (\ref{kernel}), which we would expect from the formula (\ref{mini}). Indeed, after integration, we find the following form
\beq
\boxed{
\begin{split}
     \omega(\vartheta) &= \frac{1}{2}\log 2 + \log\cosh\frac{\vartheta}{2} - \frac{i\vartheta}{2\pi} \log\left[\frac{i \cos\frac{\pi b}{2} - \sinh\vartheta}{i \cos\frac{\pi b}{2} + \sinh\vartheta}\right]  \\
     &- \frac{1}{4} \log\left[\left(\cosh\vartheta + \sin\frac{\pi b}{2}\right)\left(\cosh\vartheta - \sin\frac{\pi b}{2}\right)\right] - \frac{b}{4} \log\left[\frac{\cosh\vartheta + \sin\frac{\pi b}{2}}{\cosh\vartheta - \sin\frac{\pi b}{2}}\right]  \\
    &- \frac{i}{4\pi} \left[\mathrm{Li}_2\left( - i e^{\vartheta - i\frac{\pi}{2}b} \right) - \mathrm{Li}_2\left( i e^{\vartheta - i\frac{\pi}{2}b} \right) + \mathrm{Li}_2\left( - i e^{\vartheta + i\frac{\pi}{2}b} \right) - \mathrm{Li}_2\left( i e^{\vartheta + i\frac{\pi}{2}b} \right) + \left(\vartheta\to -\vartheta\right)\right]\,,\label{eq:f_integrated}
\end{split}
}
\eeq
where $\vartheta=i\pi-\theta$, $\mathrm{Li}_2(z)$ denotes the dilogarithm
\beq 
    \mathrm{Li}_2(z) = \sum_{k=1}^{\infty} \frac{z^k}{k^2}\;,
\eeq 
and $\frac{1}{2}\log 2$ is the integration constant that we fixed by imposing the asymptotic condition\footnote{Strictly speaking, this holds for $\mathrm{Im(\theta)}/\pi\in \mathbb{Z}$, and $|b|<1$. For other values of $\theta$ and $b$, see Appendix \ref{ApD}.}
\beq
    \lim_{|\mathrm{Re}(\vartheta)|\to\infty} \omega(\vartheta) = 0\,.
\eeq
Remarkably, we obtain a convergent formula for the logarithm of the MFF, which is explicit and contains no integrals or infinite sums. {Concerning the} numerical evaluation of the MFF, this representation is very efficient. Our numerical experiments -- performed with Mathematica on a standard laptop -- have found that a mixed representation for the MFF takes the order of $10^{-3}$ seconds to evaluate, while our formula takes the order of $10^{-4}$ seconds.  The evaluation speed largely depends on the particular implementation of the dilogarithm function used by Mathematica and may be different for other packages. In Appendix~\ref{ApD}, we comment more on the numerical evaluation, extension to values of $|b|$ in the region $1<|b|<2$, and the treatment of the function's branch cuts. We present the numerically most efficient form of \eqref{eq:f_integrated} for real $\theta$ values in Appendix~\ref{numerics}.

\begin{figure}
\begin{center}
	\includegraphics[width=12cm]{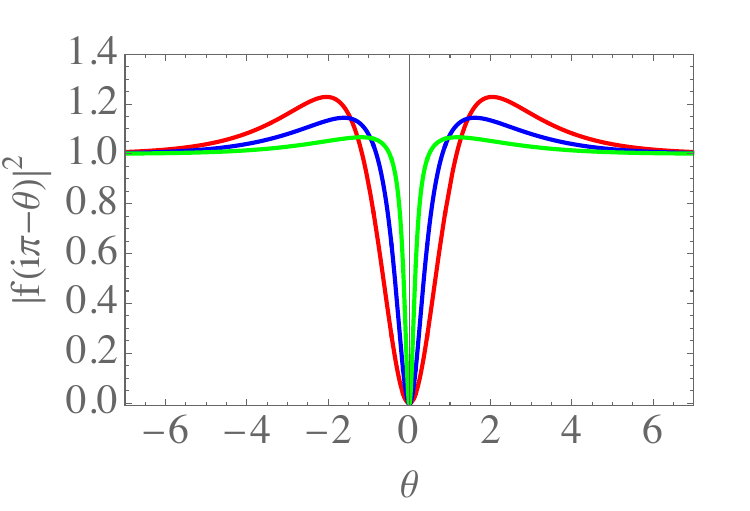}
				    \caption{The modulus square of the MFF for $b=0$ (red), $b=0.7$ (blue) and $b=0.9$ (green). Plotted with Mathematica from the formula (\ref{eq:f_integrated}).}
				     \label{fig2}
    \end{center}
    \end{figure}

The formula (\ref{eq:f_integrated}) contains several clearly identifiable contributions: 
\begin{enumerate}
\item The term 
\beq 
\log \cosh\frac{\vartheta}{2}=\log \left(-i \sinh\frac{\theta}{2}\right)\,,
\label{34}
\eeq 
contributes a factor $-i \sinh\frac{\theta}{2}$ to the minimal form factor. This results from isolating the factor $-1$ in the $S$-matrix (the Ising $S$-matrix) and produces the  MFF of the Ising field theory. 
\item The term
\beq 
- \frac{i\vartheta}{2\pi} \log\left[\frac{i \cos \frac{\pi b}{2}- \sinh\vartheta}{i \cos \frac{\pi b}{2} + \sinh\vartheta}\right]= - \frac{i (i\pi-\theta)}{2\pi} \log\left[\frac{i \cos \frac{\pi b}{2} - \sinh\theta}{i \cos \frac{\pi b}{2} + \sinh\theta}\right]=\frac{ \theta-i\pi}{2\pi} i\log (\Phi_{\bal}^{\rm shG}(\theta))\,,
\eeq 
which is precisely the structure of $\varphi_{\bal}(\theta)$ for $\bal$ chosen as in (\ref{alphas}).
\end{enumerate}
We now replace $\vartheta=i\pi-\theta$ everywhere and, with some abuse of notation, we call the logarithm of the MFF $\omega(\theta)$. We can write (\ref{eq:f_integrated}) as
\beq
    \omega(\theta) =\log\left(-i \sinh\frac{\theta}{2}\right) + \frac{ \theta-i\pi}{2\pi} i\log (\Phi_{\bal}^{\rm shG}(\theta)) +  \log(\sqrt{2} \, C^{\rm shG}_{\bel}(\theta))\,,
\label{eq:f_form}
\eeq
where 
\beqa 
 \log(C^{\rm shG}_{\bel}(\theta))= - \frac{1}{4} \log\left[\left(\cosh\vartheta + \sin\frac{\pi b}{2}\right)\left(\cosh\vartheta - \sin\frac{\pi b}{2}\right)\right] - \frac{b}{4} \log\left[\frac{\cosh\vartheta + \sin\frac{\pi b}{2}}{\cosh\vartheta - \sin\frac{\pi b}{2}}\right]\quad && \nonumber \\
 - \frac{i}{4\pi} \left[\mathrm{Li}_2\left(  i e^{-\theta - i\frac{\pi b}{2}} \right) - \mathrm{Li}_2\left( -i e^{-\theta - i\frac{\pi b}{2}} \right) +\mathrm{Li}_2\left(i e^{-\theta + i\frac{\pi b}{2}} \right) - \mathrm{Li}_2\left( -i e^{-\theta + i\frac{\pi b}{2}} \right) + \left(\theta\to -\theta\right)\right]\,.&&
\label{funC}
\eeqa 
Although it is not evident from the above expression, the function $\log(C^{\rm shG}_{\bel}(\theta))$ expands as predicted by \eqref{mini}
\beq
    \log(C^{\rm shG}_{\bel}(\theta))=\sum_{s\in\mathcal{S}'} \beta_s m^{2s}\cosh(s\theta)\;.
\eeq
The derivation of this identity and the specific $\beta_s$ coefficients are presented in Appendix~\ref{ApA}. 

\section{Minimal Form Factor of Branch Point Twist Fields}
\label{sec:minff_BPTF}
For the BPTF, the MFF has the general form (\ref{finalMFF}). In this section, we will show that this standard MFF also admits a representation of the type (\ref{mini2}). The steps are very similar to those in the previous section. However, the analytic computation of the resulting integrals is more difficult. 
Similar to (\ref{logarit}) for the BPTF, we can write 
\beq 
\omega_n(\vartheta):=\log(f_n(\vartheta))=-4\int_0^\infty \frac{dt}{t} \frac{\sinh\left(\frac{t(1+b)}{4} \right)\sinh\left(\frac{t(1-b)}{4} \right) \sinh\frac{t}{2}}{\sinh(nt)\sinh t} \cos\frac{t\vartheta}{\pi}\,,
\label{eq:logarit_n}
\eeq 
where we now call the MFF $f_n(\vartheta)$, and we recall that $\vartheta=i\pi n-\theta$. Just as in the previous section, we have that 
\beqa
\omega_n'(\vartheta)&=&\frac{1}{\pi}\int_0^\infty dt\, \frac{\sin\frac{t\vartheta}{\pi}}{\sinh(nt)}- \frac{1}{\pi} \int_0^\infty dt \, \frac{\sinh\left(\frac{t(1+b)}{2} \right)\sin\frac{t\vartheta}{\pi}}{\sinh(nt)\sinh t} - \frac{1}{\pi}\int_0^\infty dt \, \frac{\sinh\left(\frac{t(1-b)}{2} \right)\sin\frac{t\vartheta}{\pi}}{\sinh(nt)\sinh t} \nonumber\\
&=& \frac{1}{\pi}\left[g_n(\vartheta)-h_n(\vartheta;b)-h_n(\vartheta;-b)\right]\,.
\eeqa 
We now evaluate these integrals explicitly.  We find\footnote{This is identity BI (264)(16) in \cite{gr}, page 509.}
\beq 
g_n(\vartheta)=\frac{\pi}{2n}\tanh\frac{\vartheta}{2n}\,.
\label{24}
\eeq 
However, the integral $h_n(\vartheta,b)$ for $n\neq 1$ is considerably more involved than the case $n=1$. Using standard methods to compute it (such as expanding the $\sinh t$ in the denominator of the integrand), we immediately find ourselves with an infinite sum of $\Gamma$-functions, one of the standard representations of the MFF. Alternatively, we can employ contour integration as follows. 

Consider again the integral 
\beq 
h_n(\vartheta;b)=\frac{1}{2} \int_{-\infty}^\infty  dt\, \frac{\sinh\frac{t (1+b)}{2} \sin \frac{t \vartheta}{\pi}}{\sinh(nt) \sinh{t}}\,,
\label{intin}
\eeq 
where, w.r.t. the original definition,  we just doubled the integration region. We introduce the integral
\beq 
I(A,B):= \int_{-\infty}^\infty  dt F(A,B,t)\,,\quad \mathrm{with}\quad
F(A,B,t)=\frac{e^{B t} e^{i A t}}{\sinh(nt) \sinh t}\,.
\label{IAB}
\eeq 
Clearly, (\ref{intin}) is reproduced by the following sum 
\beq 
h_n(\vartheta;b)=\frac{1}{8i}\left(I\left(\frac{\vartheta}{\pi},\frac{1+b}{2}\right)-I\left(-\frac{\vartheta}{\pi},\frac{1+b}{2}\right)-I\left(\frac{\vartheta}{\pi},-\frac{1+b}{2}\right)+I\left(-\frac{\vartheta}{\pi},-\frac{1+b}{2}\right)\right)\,.
\label{hnwithI}
\eeq 
 We will now compute (\ref{IAB}) by suitably deforming the integration contour. 
 
\begin{figure}[h!]
\begin{center}
	\includegraphics[width=10cm]{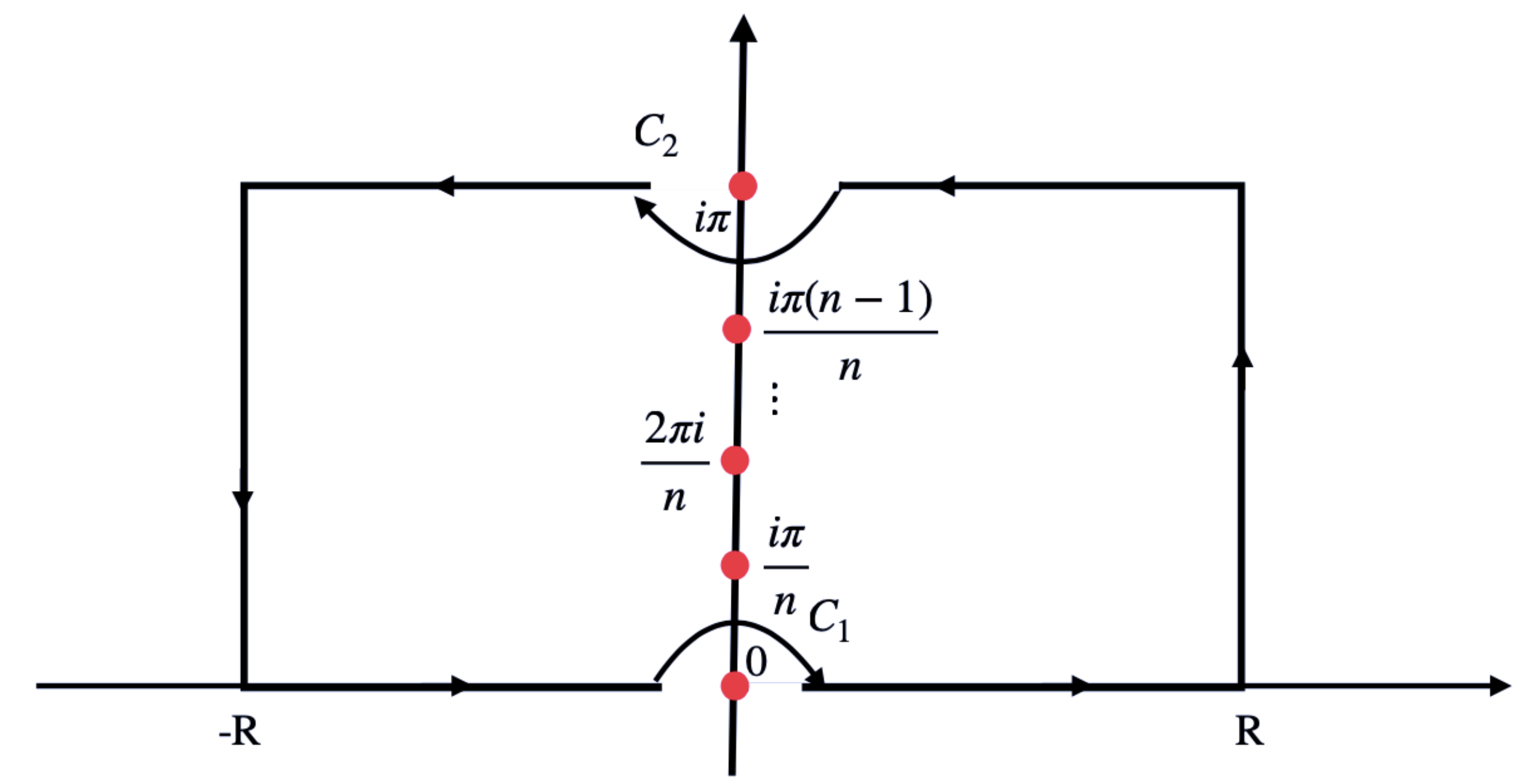}
\caption{The integration contour for integral (\ref{45}).}
				     \label{figure1}
    \end{center}
    \end{figure}

Consider a rectangular, positively oriented contour (see Fig.~\ref{figure1}) with horizontal segments at $y=0$ and $y=\pi$ and vertical segments at $x=\pm R$ with $R \rightarrow \infty$. The integrand $F(A,B,t)$ has double poles at $z=0, i\pi$ and simple poles inside the contour at $z=\frac{i\pi k}{n}$ with $k=1,\dots,\lceil n \rceil-1$ where $\lceil n \rceil$ is the ceiling function. Our integral $I(A,B)$ can then be obtained by writing
\beqa 
\oint dz\, F(A,B,z) &=& I(A,B)+ (-1)^{n} e^{i\pi B-A \pi} I(A,B) + \left[\int_{C_1}+\int_{C_2}\right] F(A,B,z) dz \nonumber\\
&=& 2\pi i \sum_{k=1}^{\lceil n \rceil-1}\mathrm{Res}_{z=\frac{i\pi k}{n}} F(A,B,z)\,,
\label{45}
\eeqa 
where the contours $C_1, C_2$ are semicircles above $z=0$ and below $z=i\pi$, and we assume -- this can be easily shown -- that the contributions of the vertical segments vanish when $R\rightarrow \infty$. We also used the fact that the contribution of the line $z\in\mathbb{R}+i\pi$ is proportional to the contribution of the integral along the real axis.
We now compute each contribution, starting with the simplest one, the residua of the simple poles at $i\pi k/n$:
\beq 
2\pi i\mathrm{Res}_{z=\frac{i\pi k}{n}} F(A,B,z)=(-1)^k\frac{2\pi e^{\frac{i\pi k}{n}(B+i A) }}{n \sin\frac{k\pi}{n}}\,.
\eeq 
The integrals over the two semicircles are also easy to evaluate as half the residue at the relevant pole (with an additional sign in the case of $C_1$ due to the orientation of the contour). This gives
\beqa 
\int_{C_1} F(A,B,z) dz=  - \frac{i\pi (B+ i A)}{n}\,,\quad \mathrm{and}\quad 
\int_{C_2} F(A,B,z) = e^{i\pi (B+i A)} (-1)^{n}   \frac{i\pi (B+ i A)}{n}\,.
\eeqa
Hence, we have
\beqa 
&& I(A,B)+ (-1)^{n} e^{i\pi B-A \pi} I(A,B) + \left[\int_{C_1}+\int_{C_2}\right] F(A,B,z) dz \nonumber\\
&&= I(A,B)\left(1+(-1)^{n}e^{i\pi(B+i A)}\right)- \frac{i\pi}{n}(B+i A)(1+(-1)^{n+1} e^{i\pi (B+iA)})\nonumber\\
&& = \sum_{k=1}^{\lceil n \rceil-1} (-1)^k\frac{2\pi e^{\frac{i\pi k}{n}(B+i A) }}{n \sin\frac{k\pi}{n}}\,.
\eeqa 
So, $I(A,B)$ receives two contributions, one from the semicircles and one from the residua. We can write
\beq 
I(A,B)=I_1(A,B)+I_2(A,B)\,,
\eeq 
with 
\beqa 
I_1(A,B):=\frac{i\pi}{n}(B+iA)\frac{1-(-1)^{n} e^{i\pi (B+iA)}}{1+(-1)^{n} e^{i\pi (B+iA)}}\,,
\eeqa 
and 
\beqa 
I_2(A,B):=\sum_{k=1}^{n-1}\frac{2\pi (-1)^k}{n \sin\frac{\pi k}{n}} \frac{e^{\frac{i\pi k}{n}(B+iA)}}{1+(-1)^{n}e^{i\pi(B+i A)}}\,.
\eeqa
With the help of Mathematica, it is relatively easy to show that 
\beq
\begin{split}
h_n(\vartheta;b)=& \frac{1}{4 n} \frac{(b+1) \pi \sinh\vartheta +(-1)^{n} 2\vartheta \cos\frac{\pi b}{2}}{\cosh\vartheta -(-1)^{n} \sin\frac{\pi b}{2}} \\
&+\frac{1}{8i}\left(I_2\left(\frac{\vartheta}{\pi},\frac{1+b}{2}\right)-I_2\left(-\frac{\vartheta}{\pi},\frac{1+b}{2}\right)-I_2\left(\frac{\vartheta}{\pi},-\frac{1+b}{2}\right)+I_2\left(-\frac{\vartheta}{\pi},-\frac{1+b}{2}\right)\right)\,,
\end{split}
\label{eq:h_n}
\eeq
and the contribution from the functions $I_1(A,B,t)$ to the sum $h_n(\vartheta;b)+h_n(\vartheta;-b)$ is
\beq 
\frac{(-1)^n 2\vartheta \cosh\frac{\pi b}{2} \cosh\vartheta+ \pi( \cosh\vartheta + (-1)^n b\sin\frac{\pi b}{2})\sinh\vartheta}{n(\cosh 2\vartheta+\cos(\pi b))}\,.
\label{cont}
\eeq 
Note that all the $(-1)^n$ terms are cancelled out if we rewrite the formulae in terms of $\theta$ rather than $\vartheta$. The integral of this term will give us a non-trivial part of the minimal form factor. In particular, the first term is proportional to (\ref{kernel}), so its integration will produce the logarithm of the $S$-matrix that we expect from (\ref{mini2}). We can also see it simplifies to part of (\ref{31}) in the $n=1$ limit. 

Putting everything together, we then have
\beq
\begin{split}
&\omega_n'(\vartheta)=\frac{1}{2n}\tanh\frac{\vartheta}{2n}-\frac{(-1)^n 2\vartheta \cosh\frac{\pi b}{2} \cosh\vartheta+ \pi( \cosh\vartheta + (-1)^n b\sin\frac{\pi b}{2})\sinh\vartheta}{\pi n(\cosh 2\vartheta+\cos(\pi b))} \\
& -\frac{1}{8 i \pi} \left(I_2\left(\frac{\vartheta}{\pi},\frac{1+b}{2}\right)-I_2\left(-\frac{\vartheta}{\pi},\frac{1+b}{2}\right)-I_2\left(\frac{\vartheta}{\pi},-\frac{1+b}{2}\right)+I_2\left(-\frac{\vartheta}{\pi},-\frac{1+b}{2}\right)+(b\mapsto -b)\right)\,.
\end{split}
\eeq

Now, we need to integrate these functions. The first line is similar to the $n=1$ case and gives a similar result. However, the contribution from $I_2(A,B)$ is unique to the BPTF and is indeed vanishing for $n=1$ (the sum over residues is not present in this case). Using Mathematica, we can show that
\beqa 
\int \left(I_2\left(\frac{\vartheta}{\pi},B\right)-I_2\left(-\frac{\vartheta}{\pi},B\right)\right) d\vartheta = \sum_{k=1}^{n-1}\frac{2\pi (-1)^k}{k \sin\frac{\pi k}{n}} \mathcal{R}(\vartheta,B,k)\,,
\eeqa 
with 
\beqa
\mathcal{R}(\vartheta,B,k)&=& e^{\frac{i\pi k B}{n}}\left[-e^{-\frac{k \vartheta}{n}} +e^{-\frac{k \vartheta}{n}} {}_2F_1(1,-\frac{k}{n},1-\frac{k}{n};(-1)^{n+1} e^{-i\pi B+\vartheta})\right.\nonumber\\
&& \qquad\qquad\qquad\quad \left. -e^{\frac{k \vartheta}{n}} {}_2F_1(1,\frac{k}{n},1+\frac{k}{n};(-1)^{n+1} e^{i\pi B+\vartheta})\right]= \mathcal{R}(-\vartheta,B,k)\,,
\label{funR}
\eeqa 
given in terms of hypergeometric functions. 
For integer $k, n$ while $k<n$, the following relations hold
\beqa
{}_2F_1\left(1,\frac{k}{n},1+\frac{k}{n};z\right)&= -\frac{k}{n} z^{-k/n} \sum_{l=0}^{n-1} e^{-2\pi i l k/n} \log(1-z^{1/n} e^{2\pi i l/n})\,, \\
{}_2F_1\left(1,-\frac{k}{n},1-\frac{k}{n};z\right)&= 1+\frac{k}{n} z^{k/n} \sum_{l=0}^{n-1} e^{2\pi i l k/n} \log(1-z^{1/n} e^{2\pi i l/n})\,.
\eeqa
Utilising this, we can simplify the combinations of the $\mathcal{R}$ terms. By introducing
\beq
\mathcal{L}(\vartheta, b, k)= \log\left[\cos\frac{b\pi}{2n}+\cos\frac{k \pi+ 2i\vartheta}{2n}\right]\,,
\eeq
the logarithm of the MFF of the BPTF takes the form
\beq
\boxed{
\begin{split}
    & \omega_n(\vartheta) = \frac{2n-1}{2n} \log 2 + \log\cosh\frac{\vartheta}{2n} + \frac{i\vartheta (-1)^n}{2\pi n} \log \Phi_{\bal}^{\rm shG}(\vartheta)  \\
    &- \frac{1}{4n} \log\left[\left(\cosh\vartheta - (-1)^n \sin\frac{\pi b}{2}\right)\left(\cosh\vartheta + (-1)^n \sin\frac{\pi b}{2}\right)\right]- \frac{b}{4n} \log\left[\frac{\cosh\vartheta - (-1)^n \sin\frac{\pi b}{2}}{\cosh\vartheta + (-1)^n \sin\frac{\pi b}{2}}\right]\\
    &+\frac{i (-1)^n}{4\pi n} \left[\mathrm{Li}_2\left( - i e^{\vartheta - i\frac{\pi}{2}b} \right) - \mathrm{Li}_2\left( i e^{\vartheta - i\frac{\pi}{2}b} \right) + \mathrm{Li}_2\left( - i e^{\vartheta + i\frac{\pi}{2}b} \right) - \mathrm{Li}_2\left( i e^{\vartheta + i\frac{\pi}{2}b} \right) + \left(\vartheta\to -\vartheta\right)\right]\\
    &-  \sum_{k=1}^{\lfloor n/2 \rfloor}\frac{n+1-2k}{2n} \left[\mathcal{L}(\vartheta, b, 3-4k)+\mathcal{L}(-\vartheta, b, 3-4k)-\mathcal{L}(\vartheta, b, 1-4k)-\mathcal{L}(-\vartheta, b, 1-4k)\right]\,, 
\end{split} }
\label{eq:f_integratedn}  
\eeq
where $\vartheta=i\pi n-\theta$.

The above expression is, once again, a new convergent and explicit representation of the MFF, which no longer involves any integrals or infinite sums\footnote{Note that all the factors $(-1)^n$ are cancelled when we write the formula in terms of the variable $
\theta$ using $\vartheta=i\pi n-\theta$. This is important if we want to analytically continue the formula for non-integer values of $n$. This can be done only after expressing everything in terms of $\theta$.}. Recall that $\Phi_{\bal}^{\rm shG}(\vartheta)=-S(\theta)$. As before, the requirement for the asymptotics fixes the integration constant. The term $ \log\cosh\frac{\vartheta}{2n}$ is the logarithm of the MFF in the Ising field theory, and the term involving $\Phi_{\bal}^{\rm shG}(\vartheta)$ is the main term in the representation (\ref{mini2}).
\begin{figure}
\begin{center}
	\includegraphics[width=7cm]{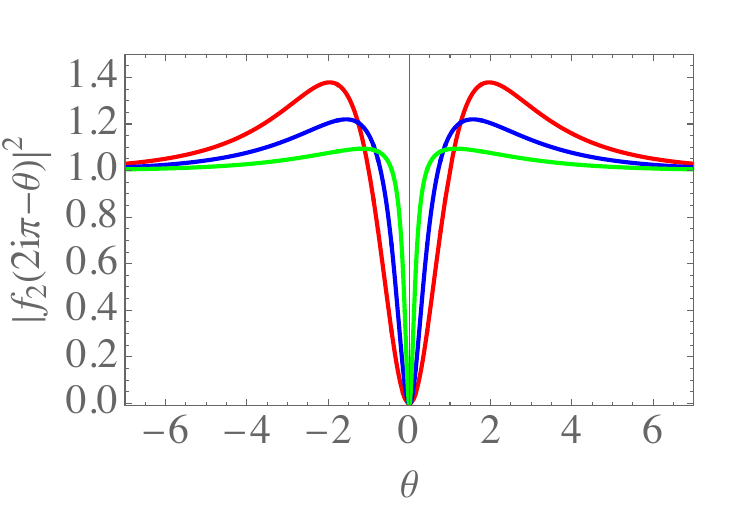}
 \includegraphics[width=7cm]{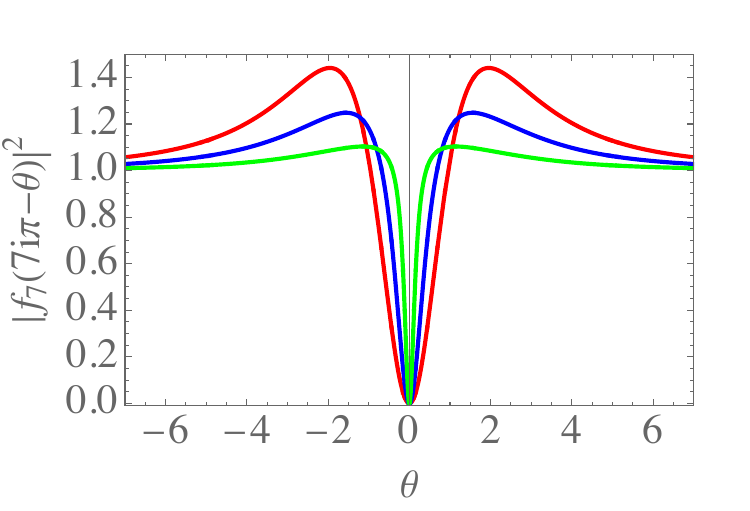}
				    \caption{The square of the MFF for $b=0$ (red), $b=0.7$ (blue) and $b=0.9$ (green) and for $n=2$ and $n=7$. Plotted with Mathematica from the formula (\ref{eq:f_integratedn}). Although both figures, as well as Fig.~2, look almost identically, they are all subtly different.}
				     \label{fig3}
    \end{center}
    \end{figure}
 As in the previous section, with some abuse of notation, we rewrite the function completely in terms of the variable $\theta$ and get
 \beq
    \omega_n(\theta) =\log\left(-i \sinh\frac{\theta}{2n}\right) + \frac{ \theta-i\pi n}{2\pi n} i\log (\Phi_{\bal}^{\rm shG}(\theta)) +  \log(2^{1-\frac{1}{2n}} \, C^{\rm shG}_{\bel}(\theta;n))\,.
\label{eq:f_formn}
\eeq
The function $C^{\rm shG}_{\bel}(\theta;n)$ can be expanded in terms of functions $\cosh\frac{s\theta}{n}$ as shown in Appendix~\ref{ApB}. The constant term in the representation (\ref{eq:f_integratedn}) is fixed by the asymptotics for $|\vartheta|$ large. As we see in Fig.~\ref{fig3}, this is such that the modulus tends to value 1 in this limit. We comment on the possible branch cuts, the extension for theories with $1<|b|<2$ and the numerical evaluation of the function in Appendix~\ref{ApD}. A slight modification of (\ref{eq:f_integratedn}) and of its $n=1$ version  (\ref{eq:f_integrated}), which is more numerically efficient by halving the number of the dilogarithm functions involved, is presented in Appendix \ref{numerics}.

\section{Important Properties of the MFF}
\label{sec:MFF_properties}

The relationship between the MFF of the sinh-Gordon model and the MFF of a generalised $\TTb$-perturbed theory that we have established in the previous two sections is instructive, especially in helping us understand the role of the functions $C_{\bel}(\theta)$ and $C_{\bel}^n(\theta)$ in the minimal form factors (\ref{mini}) and (\ref{mini2}). 

On the one hand, we know that these functions are not essential for solving the form factor equations. Indeed, in our works \cite{PRL,longpaper,entropyTTB}, we set them to $1$. We have, in fact, infinitely many minimal solutions to the two-particle form factor equations, each parameterised by a particular choice of $\bel$. In the $\TTb$ context, this is problematic since we have no physical intuition that helps us single out a specific choice of the parameters $\bel$. 

On the other hand,  we have just shown that the standard representations of MFFs (\ref{finalMFF}) correspond to a very precise, non-trivial choice of the functions $C_{\bel}(\theta)$ and $C_{\bel}^n(\theta)$. Why is this then the ``canonical" choice? A simple answer is that the integral representation satisfies the bootstrap equations in a very ``natural" way. In other words, given the $S$-matrix (\ref{S}),  the function (\ref{finalMFF}) is an obvious solution to the equations (\ref{minieq}). However, the fundamental mathematical properties that come into play are asymptotics and analyticity.

\begin{enumerate}

\item[1] {\bf Analyticity}: MFFs are minimal, meaning they only have poles on the physical strip if these are related to bound state formation, and such poles are located at specific points on the imaginary $\theta$ axis\footnote{In some conventions, the bound state singularities are factored out of the MFF, and there are no singularities in the physical strip. However, \eqref{logarit} and \eqref{eq:logarit_n} are defined in such a way as to include such singularities for the appropriate choice of the coupling $b$. This is further discussed in Appendix \ref{ApD} when examining the Lee-Yang theory.}. 
From the $S$-matrix \eqref{eq:S_matrix}, we see that in the case of sinh-Gordon theory, i.e. $0\leq |b|<1$, there are no bound state poles, but in the case of $1< |b|\leq 2$, the singularity of the $S$-matrix leads to poles at $\theta=i\frac{\pi}{2}(|b|-1)$ and $\theta= i\frac{\pi}{2} (5-|b|)$ inside the physical strip for the MFF. 

However, the singularity structure of the MFF would be even more involved if the function $C_{\bel}^{\rm shG}(\theta)$ were not present. The term $-\frac{i\vartheta}{2\pi} \log \Phi_{\bal}^{\rm shG}(\theta)$ has singularities at 
\beq 
\theta_k^{\pm}=\frac{i\pi}{2}(1\pm b)+i \pi k \,,
\eeq 
that would produce non-physical poles and zeros for the MFF inside the physical strip. The role of the logarithm contributions in $\log ( C_{\bel}^{\rm shG}(\vartheta))$ is exactly to cancel out these extra singularities. 
The logarithm terms also introduce branch cuts into $\omega(\vartheta)$, and we must be careful in defining the continuous function that reproduces the MFF. We present the details of the extra pole cancellation and the definition of $\omega(\vartheta)$ without unphysical discontinuities (see Appendix \ref{ApD} for more details). 

\item[2] {\bf Asymptotics}: The asymptotic properties of the representation (\ref{finalMFF}) are under exceptionally good control for the type of functions $g(t)$ that result from standard $S$-matrix blocks. It is easy to show that an integral of the type 
\beq 
\exp\left[2\int_0^\infty \frac{dt}{t} \frac{\cosh\left((\alpha-\frac{1}{2})t\right)}{\sinh(nt)\cosh\frac{t}{2}} \sin^2\frac{t\vartheta}{2\pi} \right]\,,
\eeq 
grows as $-i e^{\frac{\theta}{2n}}$ for $|\mathrm{Re}(\theta)|\rightarrow \infty$. This can be shown by changing variables to $x=t\theta$ and then expanding the integrand for small values of $x/\theta$ and integrating the result. For the standard minimal form factor, the same applies to $n=1$. 

In the sinh-Gordon case, we have the interplay of the exponential behaviour of two blocks and the Ising minimal form factor for large $|\vartheta|$ that produces constant asymptotics. This is a feature of the MFFs of many IQFTs. 

The contribution $-\frac{i\vartheta}{2\pi} \log \Phi_{\bal}^{\rm shG}(\vartheta)$ scales as $-\frac{|\vartheta|}{2}$ for large $|\vartheta|$, which is compensated by the Ising MFF. We can then ask what is the role played by the function $C_{\bel}^{\rm shG}(\vartheta)$ in determining this asymptotics. It is easy to show from (\ref{funC}) that
\beq 
\log(C_{\bel}^{\rm shG}(\vartheta))\approx \frac{1}{2}\log 2 \qquad \mathrm{for}\qquad |\mathrm{Re}(\vartheta)|\rightarrow \infty\,.
\eeq 
If we look in detail at how this asymptotics is achieved, we have that the two logarithmic terms in the first line of (\ref{funC}) scale as $\frac{|\vartheta|}{2}+\frac{1}{2}\log 2$ while the dilogarithms in the second line scale as $-\frac{|\vartheta|}{2}$, hence producing constant asymptotics. As we discussed in the previous point, the proper analytic behaviour of the MFF demands the presence of the logarithmic terms, but such terms on their own would alter the asymptotics of the MFF. Thus, the dilogarithm terms are crucial in compensating for this effect without introducing new singularities inside the physical strip. Note that the dialogarithm function has a  branch cut, but it is regular at the branch point ($\mathrm{Li}_2(1)=\pi^2/6$); hence it does not introduce new poles/singularities.

\item[3] {\bf Other Properties}: It is worth mentioning that one of the most useful properties of the MFF of the sinh-Gordon model, i.e.
\beq 
f(\vartheta) f(i\pi-\vartheta)= \frac{\sinh\vartheta}{\sinh
\vartheta+ i \cos \frac{\pi b}{2}}\,.
\label{eq:ft_ft+ipi}
\eeq 
This plays a central role in determining the exact form of higher particle form factors, including their asymptotic properties. Any term in $C_{\bel}^{\rm shG}(\theta)$ that involves odd couplings $\beta_{2k+1}$ does not contribute to this property simply because 
\beq 
\cosh((2 k+1) \vartheta) + \cosh((2k+1)(i\pi-\vartheta))=0\,.
\eeq 
On the contrary, the even couplings $\beta_{2k}$ all contribute and are instrumental for the property \eqref{eq:ft_ft+ipi} to hold. 
\end{enumerate}
In summary, the function $C_{\bel}^{\rm shG}(\theta)$ is crucial to ensure the analyticity and asymptotic properties of the MFF. Every part of the function plays a significant role, which in terms of the couplings $\bel$ means that all couplings, both with even and odd indexes, with their particular values as determined in Appendix \ref{ApA}, are relevant to achieving the MFF's desired properties. It also plays a role in the expression of the polynomial part (not including the poles) of higher particle form factors, but in this case, only terms associated with couplings $\beta_{2k}$ with even index play a role. This means, in particular, that the dilogarithm functions are not involved; only elementary functions contribute to determining the elementary symmetric polynomials that characterise the higher particle form factors of the sinh-Gordon model, as famously established in \cite{FMS,KK,Lukyanov:1997bp}. 

Our focus on asymptotics stems from the fact 
that it plays an important role in the convergence of correlation functions. Indeed, the growth of form factors for particular local fields is constrained by the short-distance asymptotics of two-point functions. This was established in \cite{asymptotics,delf} by showing that the form factor expansion of a two-point function $\bra \mathcal{O}(0)\mathcal{O}^\dagger(r)\ket$ can only be convergent if the form factors of $\mathcal{O}$ diverge at most as
\beq 
\lim_{|\theta_i| \rightarrow \infty} F_k^{\mathcal{O}}(\theta_1,\ldots,\theta_k) \sim  e^{y_{\mathcal{O}}|\theta_i|}\,, \quad \mathrm{with}\quad y_{\mathcal{O}}\leq \Delta_{\mathcal{O}}\,,
\label{cons}
\eeq 
where $\Delta_{\mathcal{O}}$ is the conformal dimension of the field in the underlying UV theory.

A further constraint comes from the property of cluster decomposition in momentum space \cite{DSC}, which states that 
\beq 
\lim_{\lambda \rightarrow \infty }F_p^{\mathcal{O}_1}(\theta_1+\lambda,\ldots,\theta_k+\lambda, \theta_{k+1},\ldots, \theta_p)\sim F_k^{\mathcal{O}_2}(\theta_1,\ldots,\theta_k) F_{p-k}^{\mathcal{O}_3}(\theta_{k+1},\ldots, \theta_p)\,,
\eeq 
where, in many cases, all three fields are the same, as assumed in \cite{DSC}, but not always. For example, in the Ising model, the form factors of the fields $\sigma$ and $\mu$ mix under clustering and in more complicated theories, there can be a rich structure emerging from clustering (see, for instance, \cite{CAF}).
These properties provide strict constraints that allow for the identification of the field content of the theory.

\subsection{Generalised $\TTb$ Perturbations}

Having discussed the sinh-Gordon case, seen as the Ising model perturbed by an infinite set of generalised $\TTb$ perturbations, we now consider what our conclusions say about the case of a finite number of perturbations, for example, just $\TTb$. One of the aims of this study was to get an understanding of the role played by the generic functions $C_{\bel}(\theta)$ in (\ref{mini}), and we have seen that this function plays a key role in standard IQFTs and that its main role is to ensure analyticity of the MFF. 

In the $\TTb$ case, however, the function $\log\Phi_{\bal}(\theta)$ is generally analytic since $\Phi_{\bal}(\theta)$ is a simple exponential. Thus, the role of $C_{\bel}(\theta)$ must be a different one. In this case, the asymptotic of the MFF is also very different from that found for sinh-Gordon. 
As discussed in \cite{PRL,longpaper}, it is dominated by either double-exponential growth (if the coupling of the most divergent term $\alpha_s$ or $\beta_s$ is greater than zero) or double-exponential decrease when the coupling is negative. In the case of double-exponential growth, the form factor expansion of two-point functions is clearly divergent since the form factors violate the constraint (\ref{cons}). In the case of a double-exponential decrease, the constraint (\ref{cons}) is certainly respected, and therefore two-point functions scale as power-laws at short distances. 

As for clustering, the limit above $\lambda \mapsto \infty$ produces either a divergent result (for positive coupling) or zero (for negative coupling); hence cluster decomposition does not allow for the construction of new solutions from known ones. 

We may consider whether it is possible to choose $C_{\bel}(\theta)$ in such a way as to produce an MFF which tends asymptotically to a constant. While this is not possible with a finite number of $\beta$s, it may be possible if infinitely many $\beta$s are non-zero. Preliminary results suggest that, in that case, the price to pay is loss of analyticity, hence solving a problem by creating a worse one.

In summary, the role played by the function $C_{\bel}(\theta)$ is clear for a model such as sinh-Gordon and allows us to develop a new appreciation for the delicate balance of properties that are achieved by the standard integral representation of MFFs. However, for theories perturbed by a {\it finite} number of irrelevant perturbations, the physical and mathematical constraints that should be imposed on this function remain elusive for now. 

\section{Conclusion}\label{sec:7}

In this paper, we have shown that the minimal form factor, a function which plays a very important role in the computation of form factors in IQFTs via the form factor program \cite{KW,smirnovBook}, admits a new representation. This representation is inspired by the corresponding representation found for generalised $\TTb$-perturbed IQFTs and provides support for the statement that -- at least most -- IQFTs can be interpreted in two equivalent ways: as relevant massive perturbation of a conformal critical point or as theories constructed via a finely tuned infinite set of irrelevant perturbations away from a known IQFT. In fact, the sinh-Gordon example suggests that this known IQFT might always be either a free boson or a free fermion theory, depending on whether $S_{aa}(0)=\pm 1$. This is in agreement with the conclusions of \cite{Doyon:2021tzy} where this precise statement was shown in more generality. A nice discussion of how a set of irrelevant perturbations can be used to ``redirect" the RG flow to a different critical point can be found in \cite{Muss2,LeClair:2021wfd,Ahn:2022pia}.

Our study is easily generalisable to other diagonal IQFTs. In some cases, this generalisation is immediate, like for the Lee-Yang model, where the $S$-matrix is obtained by setting the sinh-Gordon coupling to a specific value $b=B-1=-\frac{5}{3}$ beyond its usual range (this introduces a pole in the $S$-matrix) but does not change the spectrum of the theory since the bound state pole is the same fundamental particle (i.e. $a+a \mapsto a$). 
Even in the sine-Gordon model, our results find a direct application since the sinh-Gordon $S$-matrix is precisely the first breather $S$-matrix after an appropriate analytic continuation of the coupling. More generally, the sinh-Gordon $S$-matrix is a standard block from which all diagonal IQFT $S$-matrices can be built. 

Our work provides an interesting new perspective on a very old problem, the computation of the MFF, and extends the connection between known IQFT $S$-matrices and the $S$-matrices of $\TTb$-perturbed theories to a connection between their respective MFFs. It also provides new insights into how the key analyticity and asymptotic properties of MFFs result from a subtle balancing act between different contributions, which are neatly separated in our $\TTb$-inspired representation. 

We find it remarkable that nearly 45 years after the work \cite{KW}, we now have a new representation of their MFF. This is given in terms of a small number of elementary and dilogarithm functions, and it involves no integrals, no infinite sums or infinite products. It is also a very numerically efficient representation, and we expect it will be useful in future numerical studies of correlation functions in IQFTs. 

\medskip
\noindent {\bf Acknowledgments:} The authors thank Benjamin Doyon, Karol Kozlowski, Michele Mazzoni, Giuseppe Mussardo, Fabio Sailis, Roberto Tateo and Alexander Zamolodchikov for useful discussions and feedback. Olalla A. Castro-Alvaredo thanks Javier Rodr\'iguez Laguna, Silvia Santalla and Germán Sierra for their enthusiasm, interest and kindness while listening to an informal presentation on this work back in July 2023 and the organisers of the ``10th Bologna Workshop on Conformal Field Theory and Integrable Models" held in Bologna in September 2023, for giving her the opportunity to present some preliminary results. 
Olalla A. Castro-Alvaredo thanks EPSRC for financial support under Small Grant EP/W007045/1. The work of Stefano Negro is supported by the NSF grant PHY-2210533 and by the Simons Collaboration on Confinement and QCD Strings. Istv\'an M. Sz\'ecs\'enyi  received support from Nordita that is supported in part by NordForsk. 

\appendix
\section{The Expansion of Function $C^{\rm shG}_{\bel}(\theta)$}
\label{ApA}
Consider the function (\ref{funC}). For simplicity, we will restrict ourselves to  $\theta \in \mathbb{R}$ and $|b|<1$. 
The second line, involving dilogarithms, admits an expansion in terms of $\cosh (s\theta)$ functions, with $s$ odd since, for any constant $a$
\beq 
\mathrm{Li}_2\left(a e^{\theta } \right) + \mathrm{Li}_2\left(a e^{-\theta } \right)= 2\sum_{k=1}^\infty \frac{a^k \cos(k \theta)}{k^2}\,, 
\eeq 
thus
\beqa 
&& \mathrm{Li}_2\left(  i e^{-\theta - i\frac{\pi b}{2}} \right) - \mathrm{Li}_2\left( -i e^{-\theta - i\frac{\pi b}{2}} \right) +\mathrm{Li}_2\left(i e^{-\theta + i\frac{\pi b}{2}} \right) - \mathrm{Li}_2\left( -i e^{-\theta + i\frac{\pi b}{2}} \right) + \left(\theta\to -\theta\right)\nonumber\\
&&= 8 i \sum_{k=1}^\infty \frac{\cos\frac{b k \pi}{2}\sin\frac{k \pi}{2}}{k^2} \cos(k\theta)=8 i \sum_{k=0}^\infty \frac{(-1)^k\cos\frac{b (2k+1) \pi}{2}}{(2k+1)^2} \cosh((2k+1)\theta)\,.
\eeqa  
As for the other functions, we obtain contributions proportional to either
\beqa
\log\left(-\cosh\theta + \sin\frac{\pi b}{2}\right)\left(-\cosh\theta - \sin\frac{\pi b}{2}\right)= 2\log(\frac{e^{|\theta|}}{2})+2\sum_{k=1}^\infty \frac{(-1)^{k+1} \cos {\pi k b}}{k} e^{-2k|\theta|}\,,
\eeqa
or 
\beqa
\log\left(\frac{-\cosh\theta + \sin\frac{\pi b}{2}}{-\cosh\theta - \sin\frac{\pi b}{2}}\right)=  4\sum_{k=0}^\infty \frac{(-1)^k \sin\frac{\pi (2k+1) b}{2}}{2k+1} e^{-(2k+1)|\theta|}\,.
\eeqa
Symmetrising in $\theta$ allows us to replace all exponentials with $\cosh$ functions. 
Putting everything together, we then find:
\beqa 
\log(C^{\rm shG}_{\bel}(\theta))&=& \frac{1}{2}\log 2  -\frac{1}{2}\sum_{k=1}^\infty \frac{(-1)^{k+1} \cos {\pi k b}}{k} \cosh(2k \theta) \nonumber\\
&& -b\sum_{k=0}^\infty \frac{(-1)^k \sin\frac{\pi (2k+1) b}{2}}{2k+1} \cosh((2k+1)\theta)\nonumber\\
&& +\frac{2}{\pi} \sum_{k=0}^\infty \frac{(-1)^k\cos\frac{b (2k+1) \pi}{2}}{(2k+1)^2} \cosh((2k+1)\theta)\,.
\eeqa 
This result shows that the function $C_{\bel}(\theta)$ that enters the general solution (\ref{mineq}) and which we set to $1$ in our previous works \cite{PRL,longpaper,entropyTTB}, plays an important role in the usual representation of the MFF. In addition, we see that both even and odd values of spin are involved and that there are infinitely many terms.  

\section{The Expansion of Function $C_{\bel}^{\rm shG}(\theta;n)$}
\label{ApB}

We have from Section~\ref{sec:minff_BPTF} that
\beqa 
&& \log C_{\bel}^{\rm shG}(\theta;n)= \frac{1}{n} \log C_{\bel}^{\rm shG}(\theta) \\
    &&+ \frac{i}{4} \sum_{k=1}^{\lceil n \rceil-1}\frac{(-1)^k}{k \sin\frac{\pi k}{n}} \left[\mathcal{R}(\vartheta,\frac{1+b}{2},k)- \mathcal{R}(\vartheta,-\frac{1+b}{2},k)+\mathcal{R}(\vartheta,\frac{1-b}{2},k)-\mathcal{R}(\vartheta,-\frac{1-b}{2},k)\right]\,.\nonumber
\eeqa 
Consider once more $\theta \in \mathbb{R}$ and $|b|<1$.
Then, the first line admits the same expansion we found in Appendix~\ref{ApA}, whereas the second line is a combination of exponential and hypergeometric functions that we now proceed to expand. 
From the definition (\ref{funR}) we can write
\beqa
\mathcal{R}(i\pi n-\theta,B,k)&=& (-1)^k e^{\frac{i\pi k B}{n}}\left[-e^{\frac{k \theta}{n}} +e^{\frac{k \theta}{n}} {}_2F_1(1,-\frac{k}{n},1-\frac{k}{n};-e^{-i\pi B-\theta})\right.\nonumber\\
&& \qquad\qquad\qquad\quad \left. - e^{-\frac{k \theta}{n}} {}_2F_1(1,\frac{k}{n},1+\frac{k}{n};-e^{i\pi B-\theta})\right]\,.
\label{funR2}
\eeqa
and since $\mathcal{R}(i\pi n-\theta,B,k)=\mathcal{R}(i\pi n+\theta,B,k)$ (this can be checked numerically but also shown from the properties of hypergeometric functions) we can also symmetrise in $\theta$ to
\beqa
\mathcal{R}(i\pi n-\theta,B,k)&=& \frac{1}{2} (-1)^k e^{\frac{i\pi k B}{n}}\left[- 2\cosh\frac{k \theta}{n} + e^{\frac{k \theta}{n}} {}_2F_1(1,-\frac{k}{n},1-\frac{k}{n};-e^{-i\pi B-\theta})\right.\nonumber\\
&& \left. + e^{-\frac{k \theta}{n}} {}_2F_1(1,-\frac{k}{n},1-\frac{k}{n};-e^{-i\pi B+\theta}) -  e^{-\frac{k \theta}{n}} {}_2F_1(1,\frac{k}{n},1+\frac{k}{n};-e^{i\pi B-\theta})\right.\nonumber\\
&& \left.- e^{\frac{k \theta}{n}} {}_2F_1(1,\frac{k}{n},1+\frac{k}{n};-e^{i\pi B+\theta})\right]\,.
\label{funR4}
\eeqa
We have that
\beq 
{}_2F_1(1,\pm \frac{k}{n},1\pm\frac{k}{n}; x)=1+k\sum_{p=1}^{\infty} \frac{{x^p}}{k\pm p n}\,.
\eeq 
So,
\beqa
\mathcal{R}(i\pi n-\theta,B,k)&=& -(-1)^k e^{\frac{i\pi k B}{n}}\cosh\frac{k \theta}{n}+ (-1)^k e^{\frac{i\pi k B}{n}} k \sum_{p=1}^\infty \frac{(-1)^p e^{-i\pi p B }\cosh\frac{(k- p n)\theta}{n}}{k-p n}\nonumber\\
&& - (-1)^k e^{\frac{i\pi k B}{n}} k \sum_{p=1}^\infty \frac{(-1)^p e^{i\pi p B }\cosh\frac{(k+ p n)\theta}{n}}{k+p n}\,,
\label{funR3}
\eeqa
and, the combinations in (\ref{funR2}) give
\beqa
\mathcal{R}(i\pi n-\theta,B,k)-\mathcal{R}(i\pi n-\theta,-B,k)&=& -2i(-1)^k \sin\frac{\pi k B}{n}\cosh\frac{k \theta}{n}\nonumber\\
&& + 2i (-1)^k  k \sum_{p=1}^\infty \frac{(-1)^p \sin\frac{(k- p n)B}{n} \cosh\frac{(k- p n)\theta}{n}}{k-p n}\nonumber\\
&& - 2i (-1)^k  k \sum_{p=1}^\infty \frac{(-1)^p \sin\frac{(k+p n)B}{n}\cosh\frac{(k+n p)\theta}{n}}{k+p n}\,.
\label{funRdif}
\eeqa
This shows that indeed, in the case of the BPTF, the MFF does include terms of the form $\cosh\frac{s\theta}{n}$ which we can associate with fractional spin and such terms are essential in the construction of a MFF, which has the desired asymptotic properties. The presence of this type of terms was first
postulated in \cite{entropyTTB,Hou}.

\section{Minimal Form Factor Contribution for a Basic $S$-Matrix Block}
\label{apendixC}
We have already mentioned that our discussion of the sinh-Gordon model is easily generalisable to any theories with diagonal $S$-matrices. However, in order to make our results more readily usable, we present here the derivation of the minimal form factor contribution of the most basic $S$-matrix block, that is $(x)_\theta$ defined in equation (\ref{block}). Let us also recall equation (\ref{xcurnew}), which we write here again
\beq 
(x)_\vartheta= -\exp\left(-2 \int_0^\infty \frac{dt}{t} \frac{\sinh t(1+x)}{\sinh t} \sinh\frac{t\vartheta}{i\pi}\right)=\exp\left(2 \int_0^\infty \frac{dt}{t} \frac{\sinh t-\sinh t(1+x)}{\sinh t} \sinh\frac{t\vartheta}{i\pi}\right)\,.
\label{xcur2}
\eeq 
Following the construction of Section~\ref{sec:2}, we know that such a block would give rise to a corresponding minimal form factor block of the form 
\beq 
f_x(\vartheta)=\exp\left(-\int_0^\infty \frac{dt}{t} \frac{\sinh t-\sinh t(1+x)}{\sinh^2 t} \cos\frac{t\vartheta}{\pi}\right)\,.
\label{fx}
\eeq 
Proceeding as in Section 4, we call the logarithm of the function above $\omega_x(\vartheta)$ and compute its derivative
\beq 
\omega_x(\vartheta)=-\int_0^\infty \frac{dt}{t} \frac{\cos\frac{t\vartheta}{\pi}}{\sinh t} +\int_0^\infty \frac{dt}{t} \frac{\sinh t(1+x)}{\sinh^2 t} \cos\frac{t\vartheta}{\pi}\,.
\label{omegax}
\eeq 
so 
\beq 
\omega'_x(\vartheta)=\int_0^\infty \frac{dt}{\pi} \frac{\sin\frac{t\vartheta}{\pi}}{\sinh t} -\int_0^\infty \frac{dt}{\pi} \frac{\sinh t(1+x)}{\sinh^2 t} \sin\frac{t\vartheta}{\pi}\,.
\eeq 
Comparing with equations (\ref{28})-(\ref{29}) we have that 
\beq 
\omega'_x(\vartheta)=\frac{g(\vartheta)-h(\vartheta;1+2x)}{\pi}\,,
\eeq
with $g(\vartheta)$ and $h(\vartheta;1+2x)$ given by the same formulae (\ref{30}) which we recall below
\beq
    g(\vartheta) = \frac{\pi}{2} \tanh\frac{\vartheta}{2}\;,\qquad h(\vartheta;1+2x) = \frac{1}{2} \frac{(1+x) \pi \sinh\vartheta  +\vartheta \sin{\pi x}}{\cosh\vartheta + \cos {\pi x}}\;.
\eeq
The integral of $g(\vartheta)$ is simple and was already given earlier in (\ref{34}). The integral of $h(\vartheta; 1-2x)$ can also be done and expressed in terms of both elementary and dilogarithm functions
\beqa
\omega_x(\vartheta)&=&c+\log\cosh\frac{\vartheta}{2}-\frac{1+x}{2}\log(\cosh \vartheta+\cos \pi x)+i(1+2x) \tan^{-1}\left(\tan\frac{\pi x}{2}\tanh\frac{\vartheta}{2}\right)\nonumber\\
&& +\frac{x}{2}\log (x)_\vartheta - \frac{i}{2\pi}\mathrm{Li}_2\left((-x)_\vartheta e^{i x \pi}\right)+\frac{i}{2\pi} {\mathrm{Li}}_2\left(1-2i e^{-i\pi x}\frac{\sin \pi x}{1+e^{\vartheta-i\pi x}}\right)
\eeqa
In this case, we do not get a term proportional to $- i \vartheta \log (-(x)_\vartheta)$. This is because the block $(x)_\vartheta$ is not a CDD factor on its own: it needs to be combined with $(1-x)_\vartheta$ to achieve this. Nonetheless, the representation above can be used, and it is numerically efficient\footnote{Note that the overall minus sign which is present in the representation (\ref{xcur2}) and which ultimately gives rise to the contribution $g(\vartheta)$ may not be there when considering a full scattering amplitude which typically includes products of many such blocks. Whether the $\log \cosh\frac{\vartheta}{2}$ is ultimately present in the minimal form factor or not will depend on whether or not any minus signs are ``left over" when writing the full scattering amplitude.}. 

Unlike the case of the blocks $[x]_\vartheta$ which we studied in the sinh-Gordon case, the minimal form factor contribution resulting from a block $(x)_\vartheta$ does not tent asymptotically to a constant but instead decays exponentially (thus the function $\omega_x(\vartheta)$ scales linearly with $\vartheta$ for large $\vartheta$. In fact, it is not hard to prove that the linear scaling comes from the first two terms (besides the constant $c$). Indeed
\beq 
\log\cosh\frac{\vartheta}{2}-\frac{1+x}{2}\log(\cosh \vartheta+\cos \pi x)\approx -\frac{x}{2}|\vartheta| \quad \mathrm{for}\quad |\vartheta|\rightarrow \infty\,,
\eeq 
whereas the next two terms tend to a complex value which exactly compensates for the imaginary part of the dilogarithms:
\beq 
i(1+2x) \tan^{-1}\left(\tan\frac{\pi x}{2}\tanh\frac{\vartheta}{2}\right)+\frac{x}{2}\log(x)_\vartheta \approx i(1+x) \frac{\pi x}{2} \quad \mathrm{for}\quad |\vartheta|\rightarrow \infty\,.
\eeq 
Finally, the dilogarithms scale as
\beqa 
&& -\frac{i}{2\pi}\mathrm{Li}_2\left((-x)_\vartheta e^{i x \pi}\right)+\frac{i}{2\pi} {\mathrm{Li}}_2\left(1-2i e^{-i\pi x}\frac{\sin \pi x}{1+e^{\vartheta-i\pi x}}\right) \approx -\frac{i}{2\pi}\mathrm{Li}_2\left( e^{2i x \pi}\right)+\frac{i}{2\pi} {\mathrm{Li}}_2\left(1\right)\nonumber\\
&& \qquad = -\frac{i}{2\pi}\mathrm{Li}_2\left( e^{2i x \pi}\right)+\frac{i\pi}{12}\quad \mathrm{for}\quad |\vartheta|\rightarrow \infty\,.
\eeqa 
We can fix the constant $c$ by requiring a certain asymptotic behaviour. For example, we may require
\beq 
\omega_x(\vartheta)\approx -\frac{x}{2}|\vartheta|\qquad \mathrm{for} \quad |\vartheta|\rightarrow \infty,
\eeq
in which case, we simply need to take
\beq 
c=\frac{i}{2\pi}\mathrm{Li}_2\left( e^{2i x \pi}\right)-\frac{i\pi}{12}-i(1+x)\frac{\pi x}{2}=-\frac{\mathrm{Li}_2(e^{-2\pi i x})-\mathrm{Li}_2(e^{2\pi i x})}{4\pi i}\,.
\eeq 

\section{Analytic Structure}
\label{ApD}

\subsection{Cancellation of  Unphysical Poles on the Physical Strip}

As mentioned in Section \ref{sec:MFF_properties}, the MFF should only have physically motivated poles inside the physical strip, i.e. $0\leq\mathrm{Im}(\theta)\leq 2\pi n$. The logarithmic terms in \eqref{eq:f_integratedn} can have singularities on the imaginary axis at positions 
\beq
\theta^{\pm}_k=i\frac{\pi}{2}(1\pm b)+i\pi k\,,
\eeq
where $k\in \mathbb{Z}$. This leads to the behaviour 
\beq
\exp\left(\omega_n(\theta) \right)\sim (\theta-\theta_k^{\pm})^{\delta_k^{\pm}}\,.
\eeq
We want to calculate the $\delta_k^{\pm}$ exponent that belongs to possible poles inside the physical strip and show the cancellation of the non-physical poles.

The dilogarithm functions do not have singularities, while the $\log\left[-i\sinh\frac{\theta}{2n}\right]$ term in \eqref{eq:f_integratedn} has only singularities to satisfy the exchange axiom \eqref{minieq2} for $\theta=0$, i.e. $f_n(0)=f_n(i 2 n\pi )=0$, but are regular around $\theta_k^{\pm}$. Three logarithmic terms in \eqref{eq:f_integratedn} are singular at every $\theta_k^{\pm}$ point with the behaviour
\beqa
\exp\left(- \frac{1}{4n} \log\left[\left(\cosh\theta -  \sin\frac{\pi b}{2}\right)\left(\cosh\theta +  \sin\frac{\pi b}{2}\right)\right]\right)&\sim & (\theta-\theta_k^{\pm})^{-\frac{1}{4n}} \,, \\
\exp\left(  - \frac{b}{4n} \log\left[\frac{\cosh\theta - \sin\frac{\pi b}{2}}{\cosh\theta + \sin\frac{\pi b}{2}}\right] \right)&\sim & (\theta-\theta_k^{\pm})^{\pm (-1)^k\frac{b}{4n}} \,, \\
\exp\left(  i\frac{\theta-i \pi n }{2\pi n} \log\left[\frac{i \cos\frac{\pi b}{2} - \sinh\theta}{i \cos\frac{\pi b}{2} + \sinh\theta}\right]\right)&\sim & (\theta-\theta_k^{\pm})^{- \frac{1 \pm b + 2 k - 2 n}{4 n}(-1)^k} \,. 
\eeqa
The logarithms of the form
\beq
\mathcal{L}(\vartheta, b, l)=\tilde{\mathcal{L}}(\theta, b, l)= \log\left[\cos\frac{b\pi}{2n}-\cos\frac{l \pi- 2i\theta}{2n}\right]\,,
\label{eq:Ltilde_def}
\eeq
have singularities at positions $\theta \sim i \frac{\pi}{2}(-l\pm b)+i 2 n \pi p $ with $p\in \mathbb{Z}$. The singular exponents are
\beqa
\exp\left(- \frac{n+1-2l}{2n} \tilde{\mathcal{L}}(\theta, b, 3-4l)\right) &\sim & (\theta-\theta_{2n-2l+1+2np}^{\pm})^{-\frac{n+1-2l}{2n}}\,,\\ 
\exp\left(- \frac{n+1-2l}{2n}\tilde{\mathcal{L}}(-\theta, b, 3-4l)\right) &\sim &  (\theta-\theta_{2l-2+2np}^{\pm})^{-\frac{n+1-2l}{2n}} \,,\\ 
\exp\left( \frac{n+1-2l}{2n} \tilde{\mathcal{L}}(\theta, b, 1-4l)\right) &\sim &(\theta-\theta_{2n-2l+2np}^{\pm})^{\frac{n+1-2l}{2n}} \,,\\ 
\exp\left(\frac{n+1-2l}{2n} \tilde{\mathcal{L}}(-\theta, b, 1-4l)\right) &\sim & (\theta-\theta_{2l-1+2np}^{\pm})^{\frac{n+1-2l}{2n}} \,,
\eeqa
where $l=1,2,\dots \lfloor n/2 \rfloor$.

Let us focus on the singular behaviour around $\theta_k^{\pm}$ for $k=0,1,\dots, 2n-1$, that means $p=0$ for the $\tilde{\mathcal{L}}$ terms. Combining all the exponents leads to 
\beq
\delta^{\pm}_k=0\,,
\eeq
for $k=0,1,\dots, 2n-1$. One subtlety is the case when $n$ is odd, since there is no contribution from the $\tilde{\mathcal{L}}$ terms to $\delta_{n}^{\pm}$ and $\delta_{n+1}^{\pm}$; however, they are zero nonetheless. 

There are two more exponents we want to calculate, namely $\delta_{-1}^+$ and $\delta_{2n}^-$. They both get contributions from the first three logarithms and also from  $\tilde{\mathcal{L}}(-\theta, b, -1)$ or $\tilde{\mathcal{L}}(\theta, b, -1)$. We can show that
\beq
\delta_{-1}^+=\delta_{2n}^{-}=-1\,.
\eeq

Now we have all the ingredients to discuss the analyticity of the MFF inside the physical strip. The result depends on the coupling $b$. Since the $S$-matrix \eqref{eq:S_matrix} only depends on $\cos(\pi b/2)$, it is enough to focus only on the region $0 \leq b \leq 2$ to describe the MFF. We will see that the two regions $0 \leq b < 1$ and $1 < b \leq 2$ have different analyticity structures.  For $b=1$, $\omega_n(\theta)=0$.

In the case of $0 \leq b < 1$, only the potential singularities around $\theta_k^{\pm}$ for $k=0,1,\dots, 2n-1$ are inside the physical strip. The analyticity of the MFF is governed by their $\delta^{\pm}_k$ exponents. As shown above, they all vanish, and the MFF is analytic inside the physical strip.

In case of $1 < b \leq 2$, the potential singularities inside the physical strip are around $\theta_k^{+}$ for $k=-1,0,\dots, 2n-2$ and $\theta_k^{-}$ for $k=1,2,\dots, 2n$. As shown above, their exponent all vanishes apart from $\delta_{-1}^+$ and $\delta_{2n}^{-}$. Their exponents signal single order poles at $\theta_{-1}^{+}$ and $\theta_{2n}^{-}$. These singularities are related to the poles of the $S$-matrix, which can be seen from the exchange axiom \eqref{minieq2}. In this way, they are acceptable physical singularities of the MFF. For example, when $b=\frac{5}{3}$ (or alternatively $b=-\frac{5}{3}$), the $S$-matrix corresponds to the Lee-Yang model, and the pole of the MFF corresponds to bound state formation. 

To summarise, all the logarithmic terms in $C^{\rm shG}_{\bel}(\theta;n)$ play a substantial part in shaping the analytic behaviour of the MFF and cancelling out the unwanted singularities of $\omega_n(\vartheta)$. This analysis also applies to the function $\omega(\vartheta)$ defined in (\ref{eq:f_integrated}) by setting $n=1$. 

\subsection{Asymptotics and Continuous Definition of $\omega_n(\theta)$}

$\omega_n(\vartheta)$ in \eqref{eq:f_integratedn} is a real and continuous function on the $\mathrm{Im}(\vartheta)=0$ line, i.e. $\mathrm{Im}(\theta)=  n \pi$, where it coincides with the integral representation \eqref{eq:logarit_n}. However, the logarithm and dilogarithm functions have branch cuts on the complex rapidity plane. In our formula for $\omega_n(\vartheta)$ we take the principal branch of all functions involved. Therefore, we need to be careful while continuing $\omega_n(\theta)$ away from the $\mathrm{Im}(\vartheta)=0$ line, which generally involves the inclusion of additional ``correction" terms. 

We choose the branch cuts canonically. $\log(z)$ has a branch cuts running along the negative real axis, i.e. $z\in(-\infty, 0]$, and its discontinuity is an integer multiple of $ 2\pi i$.  For the dilogarithm, $\mathrm{Li}_2(z)$, the branch cut runs along $z\in [1,\infty )$, and the discontinuity is proportional to $2\pi i\log(z)$. As a consequence, \eqref{eq:f_integratedn} has branch cuts originating at the imaginary axis at points $\mathrm{Im}(\theta)=k\frac{\pi}{2}$, and $\mathrm{Im}(z)=\pm\frac{\pi}{2}(1\pm b)+k\pi$ with $k\in\mathbb{Z}$, and run parallel to the real axis, i.e. $\mathrm{Re}(\theta)\in (-\infty,0]$ or $\mathrm{Re}(\theta)\in [0,\infty)$.

The correction terms to compensate for the branch cuts are also important to ensure the desired asymptotic behaviour 
\beq
    \lim_{|\mathrm{Re}(\theta)|\to\infty} \omega_n(\theta) = 0\,,
    \label{eq:aymp}
\eeq
for all values of $\mathrm{Im}(\theta)$.

Let us examine the large $|\mathrm{Re}(\theta)|$ limit of $\omega_n(\theta)$ term by term. The limits depend on the value of $\mathrm{Im}(\theta)$, $b$, and $\mathrm{Re}(\theta)$. We find the following behaviours 
\beq
\begin{split}
\log\left[-i\sinh\frac{\theta}{2n}\right]&\sim  -\log(2)+\frac{|\mathrm{Re}(\theta) |}{2n} +i \pi \eta_{\mathrm{Re}} \\
& \quad \times\left(\frac{\left[\left(\frac{\mathrm{Im}(\theta)}{\pi}+n\right) \mod{4 n}\right]}{2n} -1 \right)\,,  \\
- \frac{1}{4n} \log\left[\left(\cosh\theta -  \sin\frac{\pi b}{2}\right)\left(\cosh\theta +  \sin\frac{\pi b}{2}\right)\right] &\sim \frac{\log(2)}{2n}-\frac{|\mathrm{Re}(\theta) |}{2n} - \frac{i \pi \eta_{\mathrm{Re}}}{2 n} \\
& \quad \times\left(\left[\left(\frac{\mathrm{Im}(\theta)}{\pi}-\frac{1}{2}\right) \mod{1}\right] -\frac{1}{2} \right)\,, \\
 - \frac{b}{4n} \log\left[\frac{\cosh\theta - \sin\frac{\pi b}{2}}{\cosh\theta + \sin\frac{\pi b}{2}}\right]&\sim  0 \,, \\
i\frac{\theta-i \pi n }{2\pi n} \log\left[\frac{i \cos\frac{\pi b}{2} - \sinh\theta}{i \cos\frac{\pi b}{2} + \sinh\theta}\right]&\sim  -\frac{|\mathrm{Re}(\theta)| +(\mathrm{Im}(\theta)-i\pi n) \eta_{\mathrm{Re}}  }{2 n}\eta_b \eta_{\mathrm{Im}} \,, \\
\mathrm{Li}_2\left( \pm i e^{\theta + i\frac{\pi}{2}b} \right)+\mathrm{Li}_2\left( \pm i e^{-\theta + i\frac{\pi}{2}b} \right) &\sim -\frac{\pi^2}{6}-\frac{    1}{2}\Bigg(|\mathrm{Re}(\theta)| -i\pi \\
&\quad +i\pi\left[\frac{\eta_{\mathrm{Re}} \mathrm{Im}(\theta)}{\pi}+\frac{
b}{2}\pm\frac{1}{2} \mod{2}\right]  \Bigg)^2\,, \\
\tilde{\mathcal{L}}(\pm\theta, b, l)&\sim-\log(2)+\frac{|\mathrm{Re}(\theta)|}{n}- \eta_{\mathrm{Re}}i\pi \\
&\quad + \frac{\eta_{\mathrm{Re}}i\pi}{2n} \left[\frac{2\mathrm{Im}(\theta)}{\pi}\pm l  \mod{4n}\right]\,,
\end{split}
\label{eq:large_theta_limit}
\eeq
where we introduced the notation
\beq
\begin{split}
\eta_b&=\mathrm{sign}\left(\cos\frac{\pi b}{2}\right)
\,, \\
\eta_{\mathrm{Re}}&=\mathrm{sign}\left(\mathrm{Re}(\theta)\right)\,, \\
\eta_{\mathrm{Im}}&=\mathrm{sign}\left(\cos\left( \mathrm{Im}(\theta)\right)\right)\,, \\
\end{split}
\eeq
used the definition of $\tilde{\mathcal{L}}$ in \eqref{eq:Ltilde_def}, and the relation
\beq
\mathrm{Li}_2\left(z\right)+\mathrm{Li}_2\left(\frac{1}{z}\right)=-\frac{\pi^2}{6}-\frac{1}{2}\log^2(-z)\,.
\label{eq:dilog_relation}
\eeq
We can write the asymptotics of all the logarithm and dilogarithm terms in \eqref{eq:f_integratedn} as
\beq
- \frac{2n-1}{2n} \log(2)-K(\theta,n,b)\,,
\eeq
where the function $K(\theta,n,b)$ contains all the dependence on $b$ and $\theta$. 

The full formula for $K(\theta,n,b)$ is straightforward to write down from \eqref{eq:f_integratedn} and \eqref{eq:large_theta_limit}, however, it is complicated and not enlightening. We omit its presentation here and will only show its simplified form for the parameter range that is relevant to our purposes. 

From the structure of \eqref{eq:f_integratedn} and \eqref{eq:large_theta_limit} we see that 
\beq
K(\theta,n,b)=K(\theta,n,-b)\,,\qquad \mathrm{and}\qquad   K(\theta,n,b)=K(\theta,n,b+4)\,,
\eeq
and we can restrict $b$ to the two regions, i.e. $0 \leq b < 1$ and $1 < b \leq 2$ to describe the whole $K(\theta,n,b)$ function. Furthermore, we focus only on the range $-\pi<\mathrm{Im}(\theta)<2n\pi$ since it contains the physical strip and the region that can be relevant for the bound state bootstrap equation in case of appropriate $b$ value. 

After some simplification, in region $0\leq b<1$ we get
\beq
\begin{split}
\tilde{K}(\theta,n,b)&=\sum_{k=-1}^{2n-1}(-1)^k \eta_{\mathrm{Re}}  \\
&\quad \times \left[\Theta\left(\frac{\mathrm{Im}(\theta)}{\pi}-k -\frac{1-b}{2} \right)\Theta\left(k+\frac{1}{2}-\frac{\mathrm{Im}(\theta)}{\pi}\right)\left(\frac{\theta}{2 n}-\frac{i\pi}{2n}\left(k+\frac{1-b}{2}\right)  \right)\right. \\
& \quad+ \left.\Theta\left(k+1-\frac{1-b}{2}-\frac{\mathrm{Im}(\theta)}{\pi}\right)\Theta\left(\frac{\mathrm{Im}(\theta)}{\pi}-k-\frac{1}{2}\right)\left(\frac{-\theta}{2 n}+\frac{i\pi}{2n}\left(k+1-\frac{1-b}{2}\right)\right)\right] \\
&\quad+\eta_{\mathrm{Re}} 2 \pi i \Theta\left(\frac{1}{4}-\left(\frac{\mathrm{Im}(\theta)}{\pi}+1\right)^2\right)\,,
\end{split}
\eeq
while for the region $1<b\leq2$, the result is
\beq
\begin{split}
\tilde{K}(\theta,n,b)&=\sum_{k=-1}^{2n-1}(-1)^k \eta_{\mathrm{Re}}  \\
&\quad \times \left[\Theta\left(\frac{\mathrm{Im}(\theta)}{\pi}-k -\frac{b-1}{2} \right)\Theta\left(k+\frac{1}{2}-\frac{\mathrm{Im}(\theta)}{\pi}\right)\left(\frac{-\theta}{2 n}+\frac{i\pi}{2n}\left(k+\frac{b-1}{2}\right)  \right)\right. \\
& \quad+ \left.\Theta\left(k+1-\frac{b-1}{2}-\frac{\mathrm{Im}(\theta)}{\pi}\right)\Theta\left(\frac{\mathrm{Im}(\theta)}{\pi}-k-\frac{1}{2}\right)\left(\frac{\theta}{2 n}+\frac{i\pi}{2n}\left(-k-1+\frac{b-1}{2}\right)\right)\right]\\
&\quad+\sum_{k=0}^{2n}(-1)^k \eta_{\mathrm{Re}} \frac{i\pi(n-k)}{n} \Theta\left(\frac{1}{4}-\left(\frac{\mathrm{Im}(\theta)}{\pi}-k\right)^2\right)\\
&\quad+\eta_{\mathrm{Re}}  \pi i \frac{n-1}{n} \Theta\left(\frac{1}{4}-\left(\frac{\mathrm{Im}(\theta)}{\pi}+1\right)^2\right)\,,
\end{split}
\eeq
where $\Theta(x)$ denotes the Heaviside step-function and the $\sim$ on  $\tilde{K}$ indicates it is only valid in the $-\pi<\mathrm{Im}(\theta)<2n\pi$ strip. With this definition, the continuous definition of $\omega_n(\theta)$ on this strip becomes
\beq
\begin{split}
    & \omega_n(\theta) = \frac{2n-1}{2n} \log 2 + \log\left[-i\sinh\frac{\theta}{2n}\right] - \frac{1}{4n} \log\left[\left(\cosh\theta -  \sin\frac{\pi b}{2}\right)\left(\cosh\theta +  \sin\frac{\pi b}{2}\right)\right]  \\
    & - \frac{b}{4n} \log\left[\frac{\cosh\theta - \sin\frac{\pi b}{2}}{\cosh\theta + \sin\frac{\pi b}{2}}\right] + i\frac{\theta-i \pi n }{2\pi n} \log\left[\frac{i \cos\frac{\pi b}{2} - \sinh\theta}{i \cos\frac{\pi b}{2} + \sinh\theta}\right]   \\
    &+\frac{i}{4\pi n} \left[\mathrm{Li}_2\left( - i e^{\theta - i\frac{\pi}{2}b} \right) - \mathrm{Li}_2\left( i e^{\theta - i\frac{\pi}{2}b} \right) + \mathrm{Li}_2\left( - i e^{\theta + i\frac{\pi}{2}b} \right) - \mathrm{Li}_2\left( i e^{\theta + i\frac{\pi}{2}b} \right) + \left(\theta\to -\theta\right)\right]\\
    & - \frac{1}{2n} \sum_{k=1}^{\lfloor n/2 \rfloor}(n+1-2k) \left[\tilde{\mathcal{L}}(\theta, b, 3-4k)+\tilde{\mathcal{L}}(-\theta, b, 3-4k)-\tilde{\mathcal{L}}(\theta, b, 1-4k)-\tilde{\mathcal{L}}(-\theta, b, 1-4k)\right] \\
    &+K(\theta,n,b)\,,
\end{split} 
\label{eq:w_n_cont}
\eeq
which has the desired asymptotics \eqref{eq:aymp} and reproduces perfectly the integral representation \eqref{eq:logarit_n}. The results for $\omega(\theta)$ trivially follow by setting $n=1$.

We note that the expression $\omega_n(\theta)$ might have short branch cuts along segments of the imaginary axis, i.e. where the imaginary part of $\omega_n(\theta)$ jumps while crossing the imaginary axis. These cuts also appear in the integral representation but do not produce cuts for the MFF \eqref{finalMFF}; rather, they fix its complex phase.

\section{The Most Numerically Effective Representation}
\label{numerics}

In terms of numerical evaluation, the most time-consuming part of our formulae is the evaluation of the eight dilogarithm functions in \eqref{eq:w_n_cont}. This number can be easily halved by using the relation \eqref{eq:dilog_relation}. If we further assume $\theta \in \mathbb{R}$, we can write our formula in terms of the variable $\theta$ as
\beq
\boxed{
\begin{split}
    & \omega_n^{\mathrm{Im}(\theta)=0}(\vartheta) = \frac{2n-1}{2n} \log 2 + \log\left[-i\sinh\frac{\theta}{2n}\right] - \frac{1}{4n} \log\left[\left(\cosh\theta -  \sin\frac{\pi b}{2}\right)\left(\cosh\theta +  \sin\frac{\pi b}{2}\right)\right]  \\
    & - \frac{b}{4n} \log\left[\frac{\cosh\theta - \sin\frac{\pi b}{2}}{\cosh\theta + \sin\frac{\pi b}{2}}\right] + i\frac{\theta-i \pi n }{2\pi n} \log\left[\frac{i \cos\frac{\pi b}{2} - \sinh\theta}{i \cos\frac{\pi b}{2} + \sinh\theta}\right]   \\
    &+\frac{i }{2\pi n} \left[\mathrm{Li}_2\left( - i e^{\theta + i\frac{\pi}{2}b} \right)+\mathrm{Li}_2\left( - i e^{\theta - i\frac{\pi}{2}b} \right)-\mathrm{Li}_2\left(  i e^{\theta + i\frac{\pi}{2}b} \right)-\mathrm{Li}_2\left(  i e^{\theta - i\frac{\pi}{2}b} \right)+\eta_b i\pi \theta\right]\\
    & - \frac{1}{2n} \sum_{k=1}^{\lfloor n/2 \rfloor}(n+1-2k)   
    \log\left[\frac{\left(\cos\frac{b\pi}{2n}-\cos\frac{(3-4k) \pi+ 2i\theta}{2n}\right)\left(\cos\frac{b\pi}{2n}-\cos\frac{(3-4k) \pi- 2i\theta}{2n}\right)}{\left(\cos\frac{b\pi}{2n}-\cos\frac{(1-4k) \pi+ 2i\theta}{2n}\right)\left(\cos\frac{b\pi}{2n}-\cos\frac{(1-4k) \pi- 2i\theta}{2n}\right)}\right] \\
    & +\frac{1-\eta_b}{2} \eta_{\mathrm{Re}} i\pi \,.
\end{split} }
\label{eq:w_n_real_axis}
\eeq
where 
\beq
\eta_b=\mathrm{sign}\left(\cos\frac{\pi b}{2}\right)= 
\begin{cases}
    +1 & 0\leq b<1 \\
    -1 & 1\leq b<2 
\end{cases}  \,,
\eeq
and we recall that $\eta_{\rm Re}=\rm{sign}({\rm Re}(\theta))$.

\bibliography{bibliography}
\end{document}